\documentclass[11pt,a4paper,3p,times,final]{elsarticle}
\usepackage[utf8]{inputenc}
\usepackage{latexsym}
\usepackage{amsmath}
\usepackage{import}
\usepackage[colorlinks]{hyperref}

\graphicspath{{figures/},{graphs/}}

\newcommand{\bld}[1]{\boldsymbol{#1}}
\newcommand{\bbld}[1]{\bar{\boldsymbol{#1}}}
\newcommand{\curly}[1]{\mathcal{#1}}
\newcommand{\omgf}{\Omega_{\mathrm{F}}}
\newcommand{\omgfp}{\Omega_{\mathrm{P}}}
\newcommand{\pomgfp}{\partial\Omega_{\mathrm{P}}}
\newcommand{\pomgf}{\partial\Omega_{\mathrm{F}}}
\newcommand{\pomgpf}{\partial\Omega_{\mathrm{F}}^{\mathrm{p}}}
\newcommand{\pomgvf}{\partial\Omega_{\mathrm{F}}^{\mathrm{u}}}
\newcommand{\gmp}{\Gamma_{\mathrm{P}}}

\newcommand{\omgbf}{\Omega_{\Box,i}^{\mathrm{F}}}
\newcommand{\omgb}{\Omega_{\Box,i}}
\newcommand{\omgd}{\Omega_{\mathrm{D}}}
\newcommand{\gmb}{\Gamma_{\Box,j}}

\newcommand{\dx}{\,{\rm d}x}
\newcommand{\ds}{\,{\rm d}s}
\newcommand{\dw}{\bld{\delta w}}
\newcommand{\dwD}{\bld{\delta \bar{w}}}
\newcommand{\dq}{\delta q}
\newcommand{\bdq}{\delta \bar{q}}

\newcommand{\du}{\bld{u}}

\newcommand{\J}{\sqrt {J_2^D}}
\newcommand{\JJ}{J_2^D}

\newcommand{\save}[1]{\emph{#1}}

\journal{arXiv}

\begin{document}

\begin{frontmatter}



\title{Computational Homogenization of Fresh Concrete Flow Around Reinforcing Bars}

\author[label1]{F. Kola\v{r}\'{i}k\corref{cor1}}
\author[label1]{B. Patz\'{a}k}
\author[label1]{J.Zeman}
\address[label1]{Department of Mechanics, Faculty of Civil Engineering, Czech
Technical University in Prague, Th\'{a}kurova 7, 166 27, Prague; Czech Republic}


\cortext[cor1]{Corresponding author: Filip Kola\v{r}\'{i}k, Department of Mechanics, Faculty of Civil Engineering, CTU in Prague, Th\'{a}kurova 7, 166 27, Prague; CZ, phone: +420 2-2435-5417, e-mail: kolarfil@cml.fsv.cvut.cz}


\begin{abstract}
Motivated by casting of fresh concrete in reinforced concrete structures, we
introduce a numerical model of a steady-state non-Newtonian fluid flow through a
porous domain. Our approach combines homogenization techniques to represent the
reinforced domain by the Darcy law with an interfacial coupling of the Stokes
and Darcy flows through the Beavers-Joseph-Saffman conditions. The ensuing
two-scale problem is solved by the Finite Element Method with consistent
linearization and the results obtained from the homogenization approach are
verified against fully resolved direct numerical simulations.
\end{abstract}


\end{frontmatter}

\section{Introduction}

The motivation for this work comes from the computational modeling of
\emph{self-compacting concrete}~(SCC) --- a type of high-performance concrete
developed in the late nineties in Japan to produce more durable 
structures~\cite{Okamura:2003:SCC}. The increased performance is ensured by the
fact that concrete casting is driven purely by self-weight without the need for
vibration casting, which is convenient especially for highly reinforced
structures with limited space between reinforcing bars~\cite{Domone:2006:SCC}.
In comparison to the conventional concretes that are designed primarily for
their compressive strength, SCCs must meet additional rheological requirements, such
as higher liquidity, in order to ensure that the mix fills the whole form-work at
low risk of phase segregation. For this reason, the focus of the numerical
modeling of SCC is not only on the structural, but also on the casting
performance, and thus it relies on techniques of computational fluid mechanics.

Depending on the level of detail, different phenomena can be taken into
consideration when modeling fresh concrete flow,
e.g.~\cite{RousselatAl,Gram:2010:NSF,Roussel:2014:SFCC}. In the most realistic
case, fresh concrete is considered as a suspension of \emph{interacting
particles convected by a fluid}. These models can be treated numerically by
discrete particle schemes, such as the discrete element
method~\cite{Mechtcherine:2013:SFC} and smoothed particle
hydrodynamics~\cite{Deeb:2014:3DI,Deeb:2014:3DII}, or by fluid solvers coupled
to particle-tracking algorithms~\cite{Svec:2012:FSF}. However, the major
disadvantage of such simulation tools is their applicability only to material-
or laboratory-scale tests, due to computational demands of the detailed
resolution.

The constitutive models aiming at structural-scale applications consider
concrete as a \emph{homogeneous} non-Newtonian \emph{fluid}, whose rheological
properties are derived from the mix composition,
e.g.~\cite{Ferraris:2001:FCR,Banfill:2006:RFC,Mahmoodzadeh:2013:RMP}. The
concrete flow can be then efficiently simulated using the Finite Element
Method~(FEM) in the Lagrangian~\cite{Dufour:2005:NMC,Cremonesi:2010:SFF} or in
the Eulerian~\cite{Patzak:2009:MFC} setting. Of course, this efficiency comes at
the cost of a coarser description of the flow. Consequently, sub-scale
phenomena can only be accounted for approximately by post-processing simulation results,
e.g., to determine the distribution and orientation of reinforcing
fibers~\cite{Kolarik:2015:MFO}, or by heuristic modification of constitutive
parameters, e.g. to account for the effect of traditional
reinforcement~\cite{Vasilic:2011:FFC}. Especially the latter aspect is critical
in the modeling of casting processes in highly-reinforced structures, which
represent the major field of application for SCC.

In this paper, we propose an efficient approach which incorporates the effects
of traditional reinforcement on fresh concrete flow. The tools of computational
homogenization, e.g.~\cite{Michel:1999:EPC,Kanoute:2009:MMC,Geers:2010:MSC},
will be utilized to avoid the need to resolve flows around each reinforcing
bar, which would lead to excessive simulation costs comparable to those of the
particle-based models. To this purpose, the structure is decomposed into three
parts:
\begin{itemize}
\item \emph{reinforcement-free} zone occupied by a homogeneous non-Newtonian fluid,
\item \emph{reinforced} zone where a two-scale homogenization scheme is employed, and
\item homogenization-induced \emph{interface} separating the reinforced and reinforcement-free zones.
\end{itemize}
As the first step, we restrict ourselves to \emph{steady state} flows; an
extension to the transient case will be reported separately following the
framework introduced in~\cite{Patzak:2009:MFC}.

In the reinforced domain, we will assume that the reinforcing bars are rigid,
acting as obstacles to the flow, and that their size~(micro-scale) is small
compared to a characteristic size of the structure or of the concrete
form-work~(macro-scale). It now follows from the results of mathematical
homogenization theory, namely by
Sanchez-Palencia~\cite[Chapter~7]{Sanchez-Palencia:1980:NHM},
Tartar~\cite{Tartar:1980:IFF}, and Allaire~\cite{Allaire:1989:HSF} for Newtonian
fluids and by Bourgeat and
Mikeli\'{c}~\cite{Bourgeat_1993:NHB,Bourgeat:1996:HPF} for non-Newtonian
fluids~(see also~\cite{Hornung:1997:HPM} for an overview), that the flow in
this region can be accurately approximated by a \emph{homogeneous Darcy law}.
The relation between the macro-scale pressure gradient and the seepage velocity
is defined implicitly, via a micro-scale boundary value problem that represents
a Stokes flow in the representative volume element (RVE) of the reinforcing
pattern, driven by the gradient of the macro-scale pressure. For the numerical
treatment of the ensuing two-scale model, we will rely on the
variationally-consistent approach developed recently by Sandstr\"{o}m and
Larsson~\cite{SandstromLarsson} and Sandstr\"{o}m et al~\cite{SandstromEtAl},
which combines the variational multi-scale method~\cite{HughesVMS} with 
first-order computational homogenization~\cite{Michel:1999:EPC,Geers:2010:MSC}.

As a result of the homogenization procedure, an artificial interface appears
that separates the Stokes domain from the Darcy domain. In order to couple the
flows in both domains, we will employ the \emph{Beavers-Joseph-Saffman
conditions}~\cite{BeaversSaffman,Saffman} that effectively act as frictional
conditions to the Stokes flow. The ensuing interface constants, relating the
traction vector and the relative tangential slip in velocity, can be
estimated from an auxiliary boundary value problem at the cell level, derived
for Newtonian fluids by a refined asymptotic analysis in the seminal work of
J\"{a}ger and Mikeli\'{c}~\cite{JagerMikelic2000} and verified later by direct
numerical simulations for free laminar flow by J\"{a}ger et
al.~\cite{Jager:2001:AAL} and Carraro et al.~\cite{CarraroMikelic2013}.

Following these considerations, the rest of the paper is organized as follows.
In Section~\ref{sec:formulation}, we present the development of the homogenized
model, including the variationally-consistent homogenization in the Darcy domain
and a discussion of the Stokes-Darcy coupling. The numerical aspects of the
problem are gathered in Section~\ref{sec:numerical_solution}. In
Section~\ref{NumericalExamplesSec}, we address the errors introduced by the
homogenization and the coupling procedures, by comparing the homogenized model
with fully resolved simulations. The potential of the developed model is
critically discussed in the concluding Section~\ref{sec:conclusions}, where we
highlight the need for a more refined interface description.

The novelty of this paper is twofold. We see our first contribution in the
development of a systematic procedure to incorporate the effect of reinforcement
into homogeneous models, thereby \emph{rationalizing the porous media analogy}
introduced by Vasili\'{c} et al.~\cite{Vasilic:2011:FFC} on heuristic grounds.
The second novelty lies in our treatment of the homogenized model and the
Stokes-Darcy coupling \emph{simultaneously}. Indeed, there are several studies
on the Stokes-Darcy coupling with the help of proper interface conditions,
e.g.~\cite{Discacciati:2002:NNM,UrquizaSDcoupling,KarperMardalSDcoupling}, in
which the permeability is given in advance instead of being up-scaled from the
underlying micro-structure. Other works deal with the homogenization of the
Stokes flow through the porous domain,
e.g.~\cite{SandstromEtAl,SandstromLarsson}, or consider the coupled Stokes-Darcy
flow, but do not allow for the flow between the domains,
e.g.~\cite{Jager:2001:AAL,CarraroMikelic2013}. This study seems to be the first
one considering these effects simultaneously for both Newtonian and
non-Newtonian fluids. In these aspects, the present work extends our recent
contribution~\cite{Kolarik:2015:MFC} where only linear Newton rheology was
considered and where the methodology was explained is much less detail.

\section{Formulation of the problem}\label{sec:formulation}

In this section, we present the derivation of the homogenized model, employing
the framework of the variationally consistent
homogenization~\cite{SandstromLarsson,SandstromEtAl} in the bulk and the refined
asymptotic analysis of J\"ager and Mikeli\'{c}~\cite{JagerMikelic2000} at the
internal interface. For the sake of notational simplicity, we restrict ourselves
to the two-dimensional setting shown in Fig.~\ref{perforated_scheme} and refer
the readers interested in the general case
to~\cite{SandstromLarsson,SandstromEtAl}. Specifically, the obstacles are
assumed to be arranged according to a regular grid of cells, further referred to
as Volume Elements~(RVEs), and to be located symmetrically with respect to the
center of each RVE without intersecting its boundary. Further, the
perforated domain is assumed to be fully contained within the unperformed
domain, where the external boundary conditions are imposed.

\subsection{Strong form}\label{StrongForms}

\begin{figure}[h]
\centering\includegraphics{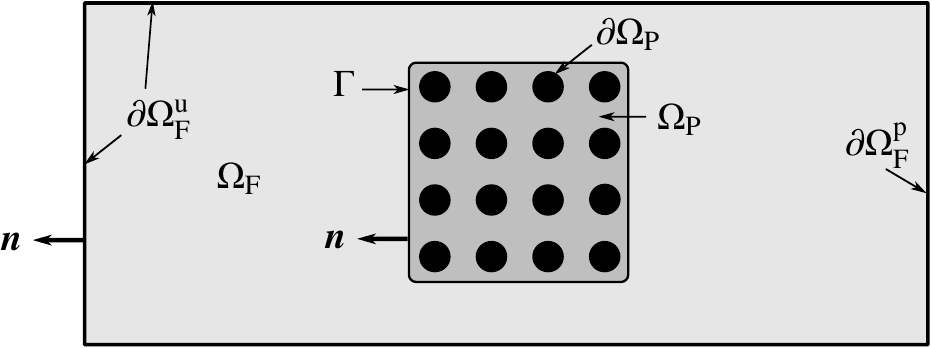} 
\caption{\textit{Stokes flow around the bars modeled as a perforated domain.}}
\label{perforated_scheme}  
\end{figure}

As a point of departure, consider the Stokes flow over a perforated domain as
shown in Fig. \ref{perforated_scheme}. We denote, in agreement with
Fig.~\ref{perforated_scheme}, $\omgf$ as the reinforcement-free part of the
domain and $\omgfp$ as a part of the domain with the obstacles (further called
perforated sub-domain). Boundary of the obstacles is denoted as $\pomgfp$, while
the outer boundary $\pomgf$ is split into two disjoint parts $\pomgpf$ and
$\pomgvf$ corresponding to the type of applied boundary condition; $\Gamma$
stands for the interface between the perforated, $\omgf$, and the unperforated,
$\omgfp$, domains. By $\bld{n}$, we denote both the outer unit normal vector to
$\pomgf$ and $\Gamma$, in the latter case pointed from to $\omgfp$ to $\omgf$.

The governing equations of the steady-state flow of an incompressible fluid in
the union of domains $\omgf\cup\omgf$ take the form
\begin{subequations}\label{eq:strong_form}
\begin{align}
-\boldsymbol{\nabla} 
\cdot 
\bld{\tau}\left( \boldsymbol{D}( \boldsymbol{u} ) \right) 
+ 
\boldsymbol{\nabla}p 
= 
\rho
\boldsymbol{b}  
&& \text{in  $\omgfp \cup \omgf$}
\label{momentum_eq}
\\
\boldsymbol{\nabla} \cdot  \boldsymbol{u} 
= 
0 
&& 
\text{in  $\omgfp \cup \omgf$}\label{incompressibility_eq}\\
  \boldsymbol{u} = \boldsymbol{0} && \text{in  $\pomgfp$}\label{BC_pomgfp}\\
  (\bld{\tau} - p\bld{I})\cdot \bld{n} = -\hat{p}\bld{n} &&\text{on $\pomgpf$}\label{BC_gmp}\\
  \boldsymbol{u} = \hat{u}_n\boldsymbol{n} && \text{on  $\pomgvf$.}\label{BC_gmv}
\end{align}
\end{subequations}
 Our notation is standard;
$\bld{\tau}$ stands for the deviatoric part of a stress tensor, the strain rate
tensor $\bld{D}$ is obtained as the symmetrized gradient of the unknown velocity
field $\bld{u}$,
\begin{equation}\label{strainrate}
  \boldsymbol{D}( \boldsymbol{u} ) = \frac{1}{2} \left(\boldsymbol{\nabla}
  \boldsymbol{u} + \left(\boldsymbol{\nabla} \boldsymbol{u} \right)^{T}
  \right), \nonumber
\end{equation}
$p$ denotes pressure, $\rho\bld{b}$ are body forces, $\bld{I}$ is the unit
second order tensor and $\hat{u}_n$ and $\hat{p}$ refer to the boundary data.
Notice that we set the velocity on the boundary of the bars $\pomgfp$ to zero
in~\eqref{BC_pomgfp}. Another physically reasonable possibility is to consider
zero velocity in the normal direction to the boundary of the bars only. However,
the former case is preferable from the numerical point of view and also it is
frequently used by others \cite{SandstromLarsson}. We believe that in the most
situations it is also more realistic, because in real castings the concrete will
stick to the bars during the flow.

\subsection{Variational form and two-scale decomposition}\label{sec:vartional}

The weak form of~\eqref{eq:strong_form} amounts to finding a pair
$(\boldsymbol{u}, p)$ such that
\begin{subequations}\label{eq:weak_stokes}
\begin{align}
  &\int_{\omgfp} \bld{\nabla} \dw : \bld{\tau}(\bld{D}(\bld{u})) \dx -
  \int_{\omgfp}( \bld{\nabla}\cdot\dw)p \dx+ \int_{\omgfp}\dq (\bld{\nabla}
  \cdot \bld{u}) \dx \label{weak_stokes1}\\ &  +\int_{\omgf} \bld{\nabla} \dw : \bld{\tau}(\bld{D}(\bld{u})) \dx - \int_{\omgf}( \bld{\nabla}\cdot\dw)p \dx+ \int_{\omgf}\dq (\bld{\nabla} \cdot \bld{u}) \dx\label{weak_stokes2} \\ &+ \int_{\pomgf} \dw \cdot\hat{p}\bld{n} \ds +\int_\Gamma \llbracket \dw\cdot\bld{\tau}\cdot\bld{n}\rrbracket\ds - \int_\Gamma \llbracket \dw\cdot p\bld{n}\rrbracket\ds\label{weak_stokes3}\\ 
& = \int_{\omgfp}\dw \cdot \rho \bld{b} \dx + \int_{\omgf}\dw \cdot \rho \bld{b} \dx\label{weak_stokes4} 
\end{align}
\end{subequations}
holds for all $(\dw, \dq)$. Here, \eqref{weak_stokes1}, \eqref{weak_stokes2},
and \eqref{weak_stokes4} represent the contributions of the perforated and
unperforated domains and \eqref{weak_stokes3} appears after the integration by
parts. The velocity field $\bld{u}$ must satisfy the essential boundary
conditions~\eqref{BC_pomgfp} on $\pomgfp$ and~\eqref{BC_gmv} on $\pomgvf$, the
test field $\dw = \bld{0}$ on $\pomgfp$ and on $\pomgvf$. In addition, all the
involved fields $(\bld{u}, p, \dw, \dq)$ may be discontinuous across the
interface $\Gamma$, which gives rise to the jump terms in~\eqref{weak_stokes3}
defined as
\begin{align}
\llbracket f \rrbracket (\bld{x}) = f(\bld{x})\bigl|_{\Omega_\mathrm{F}} - f(\bld{x})\bigl|_{\Omega_\mathrm{P}}
\text{ for }
\bld{x} \in \Gamma,\nonumber
\end{align}
where ${\Omega_\mathrm{F}}$ and ${\Omega_\mathrm{P}}$ after the vertical line symbolize
that the function $f$ is evaluated at the non-perforated or
perforated side of the interface $\Gamma$, respectively.

In order to properly average the flow in the perforated domain $\omgfp$, we
follow the idea of the variational multi-scale method by Hughes et
al.~\cite{HughesVMS} and its application to porous media by Sandstr\"{o}m and
Larsson~\cite{SandstromLarsson} and Sandstr\"{o}m et al.~\cite{SandstromEtAl},
and introduce a decomposition of the unknown pressure
field $p$ and its corresponding test function $\dq$ into the macro-scale and
sub-scale parts 
\begin{align}\label{eq:pressure_decomposition}
p = p^\mathrm{M} + p^\mathrm{S}, \quad 
\dq = \dq^\mathrm{M} + \dq^\mathrm{S} \quad
\text{ in } \omgfp. 
\end{align}

We proceed as in~\cite{SandstromLarsson} by introducing the
decomposition~\eqref{eq:pressure_decomposition} into the weak Stokes
problem~\eqref{eq:weak_stokes} and by integrating the terms
containing macro-scale pressure $p^\mathrm{M}$ and its corresponding test function
$\dq^\mathrm{M}$ by parts. The corresponding parts of~\eqref{eq:weak_stokes} then
transform into finding $(\bld{u}, p^\mathrm{M}, p^\mathrm{S}) $ such that
  \begin{align}
  &\int_{\omgfp} \bld{\nabla} \dw : \bld{\tau}(\bld{D}(\bld{u})) \dx +
  \int_{\omgfp}\dw\cdot\bld{\nabla}p^\mathrm{M} \dx- \int_{\omgfp}( \bld{\nabla}\cdot\dw)p^\mathrm{S}\dx\nonumber\\ & + \int_\Gamma \dq^\mathrm{M}(\bld{u}\cdot\bld{n})\ds -
\int_{\omgfp}\bld{\nabla}\dq^\mathrm{M}\cdot\bld{u}\dx + \int_{\omgfp}\dq^\mathrm{S} \bld{\nabla} \cdot \bld{u} \dx\nonumber\\ & -\int_\Gamma \left.\dw\cdot\bld{\tau}\right|_{\Omega_\mathrm{P}}\cdot\bld{n}\ds +  \int_\Gamma \left.\dw\cdot p^\mathrm{S}\right|_{\Omega_\mathrm{P}}\bld{n}\ds \nonumber\\ 
& = \int_{\omgfp}\dw \cdot \rho \bld{b} \dx \label{separatedWeakForm2}
 \end{align}

\save{
\begin{align}
 &\int_{\omgfp} \bld{\nabla} \dw : \bld{\tau}(\bld{D}) \dx - \int_{\omgfp}( \bld{\nabla}\cdot\dw)(p^\mathrm{M} + p^\mathrm{S}) \dx+ \int_{\omgfp}(\dq^\mathrm{M} + \dq^\mathrm{S}) (\bld{\nabla} \cdot \bld{u}) \dx \nonumber\\
& -\int_\Gamma \left.\dw\cdot\bld{\tau}\right|_{\Omega_\mathrm{P}}\cdot\bld{n}\ds  
 +  \int_\Gamma \left.\dw\cdot (p^\mathrm{M} + p^\mathrm{S})\right|_{\Omega_\mathrm{P}}\bld{n}\ds\nonumber\\ 
& = \int_{\omgfp}\dw \cdot \rho \bld{b} \dx ,\label{separatedWeakForm} 
 \end{align}
} 
holds for all $(\dw, \dq^\mathrm{S}, \dq^\mathrm{M})$, where $\bld{u}$ and $\dw$ vanish on the
reinforcement boundaries~$\partial\omgfp$, recall Eq.~\eqref{BC_pomgfp}. 

\subsection{Averaging and pressure extension}

\begin{figure}[h]
\centering\includegraphics{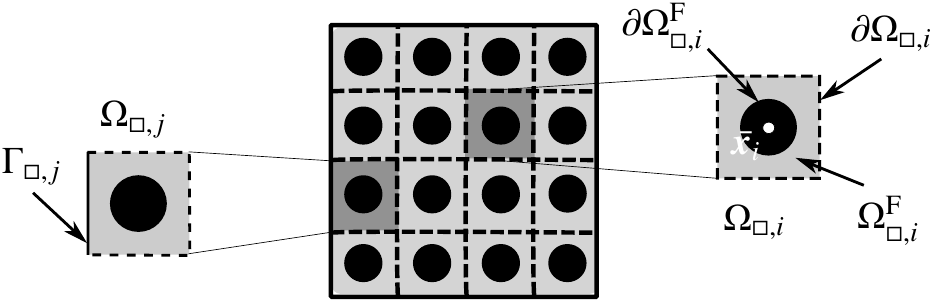} 
\caption{\textit{Decomposition of the homogenized domain $\omgd$ into the union
of Representative Volume Elements~(RVEs) $\omgb$; $\Omega_{\Box, j}$ denotes RVEs
adjacent to the fluid domain $\omgf$.}}
\label{UnionOfRVEs}  
\end{figure}

The geometry of the perforated domain, Fig.~\ref{UnionOfRVEs},  suggests that
the domain can be covered by a collection of RVEs $\Omega_{\Box,i}$
with boundary $\partial \Omega_{\Box,i}$, the part of each RVE occupied by fluid
is denoted by $\omgbf$ and $\partial \omgbf$ refers to the boundary of the bar.
The RVEs adjacent to the internal boundary $\Gamma$ are designated by the index
$j$, i.e. $\Omega_{\Box,j}$, and the corresponding parts of their boundaries are
denoted by $\Gamma_{\Box,j}$. Finally, $\omgd$ refers to the homogenized
macro-scale domain covered by RVEs.

Following~\cite{SandstromLarsson}, we define RVE-wise constant volume and
boundary averages
\begin{align}
\left < f \right >( \bld{x} ) & = \frac{1}{|\omgbf|}\int_{\omgbf}f \dx
\text{ for }
\bld{x} \in \omgb,\nonumber
\\
\left <\left < f \right >\right>( \bld{x} ) 
& = \frac{1}{|\gmb|}\int_{\gmb}f \ds
\text{ for }
\bld{x} \in \gmb,\nonumber
\end{align}
that satisfy the averaging relation, cf.~\cite[Section 3.2]{SandstromLarsson},
\begin{align}\label{averagingRelation}
\int_{\omgfp} f \dx = \int_{\omgd} \phi \left < f \right > \dx,
&& 
\int_{\Gamma} f \dx = \int_{\Gamma}\left <\left < f \right >\right> \dx,
\end{align}
where $\phi = |\omgbf| / |\omgb|$ denotes the (constant) porosity of the
perforated domain. Observe that, in contrast to~\cite{SandstromLarsson}, no coefficient
similar to $\phi$ appears in the boundary averaging, because we assume that
the obstacles do not intersect the boundary~$\Gamma$.

Using \eqref{averagingRelation}, the weak formulation
\eqref{separatedWeakForm2} can be re-defined over the homogenized domain
$\omgd$: Find $(\boldsymbol{u}, p^\mathrm{M}, p^\mathrm{S})$ such that
\begin{subequations}\label{separatedWeakForm3}
\begin{align}
&
\int_\Gamma \left<\left<\dq^\mathrm{M}(\bld{u}\cdot\bld{n})\right>\right>\ds 
-
\int_{\omgd}\phi\left<\bld{\nabla}\dq^\mathrm{M}\cdot\bld{u}\right>\dx
\label{eq:separated_weak_qM}
\\
+ & 
\int_{\omgd}\phi\left<\dq^\mathrm{S} \bld{\nabla} \cdot \bld{u}\right> \dx
+
\int_{\omgd} \phi\left<\bld{\nabla} \dw : \bld{\tau}(\bld{D}(\bld{u}))\right>
\dx 
\nonumber \\
+ & 
\int_{\omgd}\phi\left<\dw\cdot\bld{\nabla}p^\mathrm{M}\right> \dx 
- 
\int_{\omgd}\phi\left<( \bld{\nabla}\cdot\dw)p^\mathrm{S}\right>\dx
\label{eq:separated_cell_Stokes1}
\\ 
- &    
\int_\Gamma\left<\left<\dw\cdot\bld{\tau}\bigl|_{\Omega_\mathrm{P}}\cdot\bld{n}\right>\right>\ds 
+ 
\int_\Gamma \left.\left<\left< \dw\cdot p^\mathrm{S}\right|_{\Omega_\mathrm{P}}\bld{n}\right>\right>\ds
\label{eq:boundary_terms} \\ 
= &  \int_{\omgd}\phi\left<\dw \cdot \rho \bld{b}\right> \dx
\label{eq:separated_cell_Stokes2}
\end{align}
\end{subequations}
holds for all  $(\dw, \dq^\mathrm{S}, \dq^\mathrm{M})$. 

As customary in conventional homogenization theories, we assume that the RVEs
$\omgb$ are much smaller than the homogenized domain $\omgd$, so that
the contribution of the boundary terms~\eqref{eq:boundary_terms}
to~\eqref{separatedWeakForm3} is negligible. To clarify this assumption,
consider a sub-scale traction vector 
\begin{align}\label{eq:traction_def}
\bld{T}^\mathrm{S} 
= 
(\bld{\tau} - p^\mathrm{S}\bld{I}) 
\cdot \bld{n},
\end{align}
associated with the stress tensor~$\bld{\tau}$ and the sub-scale
pressure~$p^\mathrm{S}$. Neglecting~\eqref{eq:boundary_terms} thus means
neglecting the virtual work of the sub-scale tractions at the whole
boundary $\Gamma$,
\begin{align}
\int_\Gamma
\left<\left<
\dw\cdot
\bld{T}^\mathrm{S} \bigl|_{\Omega_\mathrm{P}}
\right>\right>
\ds \approx 0.\nonumber
\end{align}
We shall see later in Section~\ref{coupling} that this contribution is 
partially compensated for by the interface conditions between the
reinforcement-free and homogenized domains.

The transition from the perforated domain $\omgfp$ to the homogenized domain
$\omgd$ is completed by extending the
macro-scale pressure $p^\mathrm{M}$~(defined on $\omgfp$) to a smooth pressure
field $\bar{p}$~(defined on the whole $\omgd$), which satisfies the matching
conditions
\begin{align}
\bar{p}( \bar{\bld{x}}_i ) 
= 
\left< p^\mathrm{M} \right>( \bar{\bld{x}}_i ), 
&& 
\bld{\nabla}\bar{p}( \bar{\bld{x}}_i ) 
= 
\left< \bld{\nabla} p^\mathrm{M} \right>( \bar{\bld{x}}_i ), \nonumber
\end{align}
at the center of the $i$-th RVE $\bar{\bld{x}}_i$, recall
Fig.~\ref{UnionOfRVEs}. By analogy, we also extend the test function
$\dq^\mathrm{M}$ to $\delta\bar{q}$, from $\omgfp$ to $\omgd$. 

\subsection{First-order homogenization}

Now, we adopt the ansatz of the first-order homogenization, and expand the
unknown fields in each RVE into an affine part macro-scale and a periodic
sub-scale correction,
cf.~\cite{Larsson:2010:VCC,Sykora:2012:CHN,SandstromLarsson}. The resulting
approximations at $\bld{x} \in \Omega_{\Box,i}^\mathrm{F}$ take the form

\begin{subequations}
\begin{align} \label{pressureExpansion}
p(\bld{x}) 
& \approx 
(p^\mathrm{M} + p^\mathrm{S})(\bld{x}) 
= 
\bar{p}(\bar{\bld{x}}_i) 
+
\bld{\nabla}\bar{p}(\bar{\bld{x}}_i) 
\cdot 
(\bld{x} - \bar{\bld{x}}_i) +
p^\mathrm{S}(\bld{x}),
\\
\bld{u}(\bld{x})
& \approx 
\bld{u}^\mathrm{S}(\bld{x}), 
\label{velocityExpansion}
\end{align}
\end{subequations}
where 
\begin{subequations}\label{eq:subscale_def}
\begin{align}
p^\mathrm{S} \text{ is periodic on } \partial \omgb, 
&& 
\left<p^\mathrm{S} \right> & = 0 \text{ in } \omgb,\nonumber
\\
\bld{u}^\mathrm{S} \text{ is periodic on } \partial \omgb,
&&
\bld{u}^\mathrm{S} & = \bld{0} \text{ on } \partial \Omega_{\Box,i}^\mathrm{F}.\nonumber
\end{align}
\end{subequations}
A similar expansion is adopted for the test functions 
\begin{subequations}\label{eq:test_Expansion}
\begin{align} \label{test_pressureExpansion}
\dq(\bld{x}) 
& \approx 
(\dq^\mathrm{M} + \dq^\mathrm{S})(\bld{x}) 
= 
\delta\bar{q}(\bar{\bld{x}}_i) 
+
\bld{\nabla}\delta\bar{q}(\bar{\bld{x}}_i) 
\cdot 
(\bld{x} - \bar{\bld{x}}_i) +
\dq^\mathrm{S}(\bld{x}),
\\
\dw(\bld{x})
& \approx 
\dw^\mathrm{S}(\bld{x}), 
\text{ for }
\bld{x} \in \Omega_{\Box,i}^\mathrm{F},
\label{test_velocityExpansion}
\end{align}
\end{subequations}
with the test fields $(\dq^\mathrm{S},\dw^\mathrm{S})$ satisfying the same
conditions as the pair $(p^\mathrm{S},\bld{u}^\mathrm{S})$
in~\eqref{eq:subscale_def}.

We now proceed with testing the weak form~\eqref{separatedWeakForm3} separately
with the macro-scale, Section~\ref{sec:Darcy}, and micro-scale,
Section~\ref{sec:Stokes}, test functions, from which we extract the
corresponding macro- and micro-scale problems of the Darcy and the Stokes type,
respectively.

\subsubsection{Macro-scale Darcy law}\label{sec:Darcy}

Inserting the expansion of the macro-scale pressure test
function~\eqref{test_pressureExpansion} into~\eqref{eq:separated_weak_qM}
gives rise to the terms, cf.~\cite[Eq.~(15)]{SandstromLarsson}
\begin{align}
\int_{\omgd}\phi\left<\bld{\nabla}\dq^{\mathrm{M}}\cdot\bld{u}\right>\dx
& \approx
\int_{\omgd}\bld{\nabla}\delta{\bar{q}} \cdot \phi \left<\bld{u}\right>\dx,\nonumber
\\
\int_\Gamma \left<\left<\dq^{\mathrm{M}}(\bld{u}\cdot\bld{n})\right>\right>\ds 
& \approx
\int_\Gamma \delta
\bar{q} \left<\left<\bld{u}\right>\right> \cdot\bld{n} \ds.\nonumber
\end{align}
By introducing the seepage velocity via
\begin{align}
\bar{\bld{u}} = \phi \left<\bld{u}\right> 
\text{ in } \omgd, 
&&
\bar{\bld{u}} = \left<\left<\bld{u}\right>\right>
\text{ on } \Gamma, \nonumber
\end{align}
we finally see that~\eqref{eq:separated_weak_qM} encodes the weak
from of the mass conservation condition
\begin{align}
\int_{\omgd}\bld{\nabla} \bdq\cdot \bar{\bld{u}}\dx 
=
\int_{\Gamma}\bdq\bar{\bld{u}}\cdot\bld{n} \ds. \nonumber
\end{align}
For numerical convenience, and also for purpose of coupling the homogenized flow
in $\omgd$ to the flow in the fluid domain $\omgf$, we further express the
conservation condition in an equivalent form
\begin{align}\label{macroscaleProblem}
&\int_{\omgd} \bdq (\bld{\nabla}\cdot \bar{\bld{u}})\dx = 0,
\end{align}

To complete the macro-scale problem, in the next section we will
show that the seepage velocity is driven by macro-scale
gradient~$\bld{\nabla}\bar{p}$ and by the body forces~$\rho\bld{b}$, 
\begin{align}\label{materialLaw}
\bar{\bld{u}} = \bld{\bar{w}}(\bld{\nabla} \bar{p},\rho\bld{b})
\text{ in } \omgd,
\end{align}
so that~\eqref{macroscaleProblem} and~\eqref{materialLaw} indeed represent a
non-linear Darcian flow in $\omgd$.

\subsubsection{Micro-scale Stokes problem}\label{sec:Stokes}

The sub-scale problem is obtained from~\eqref{eq:separated_cell_Stokes1}
and~\eqref{eq:separated_cell_Stokes2} by localizing the test functions to
individual RVEs $\omgb$ according to~\eqref{eq:test_Expansion}. Abbreviating
$\bar{\bld{g}} = \bld{\nabla} \bar{p}( \bar{\bld{x}}_i )$, the resulting problem
reads as follows: Find $(\bld{u}^\mathrm{S}, p^\mathrm{S})$ such that
\begin{subequations}\label{subscaleProblem}
\begin{align}
\int_{\omgbf} 
\bld{\nabla} \dw^\mathrm{S} 
: 
\bld{\tau}(\bld{D}(\bld{u}^\mathrm{S})) 
\dx 
- 
\int_{\omgbf}( \bld{\nabla}\cdot\dw^\mathrm{S})p^\mathrm{S} \dx  & 
= 
\int_{\omgbf} 
\dw^\mathrm{S} 
\cdot 
\left( 
\rho\bld{b} - \bar{\bld{g}}
\right)
\dx, 
\label{subscaleProblem1}
\\ 
\int_{\omgbf}\dq^\mathrm{S} 
(\bld{\nabla} \cdot \bld{u}^\mathrm{S}) \dx & = 0,
\end{align}
\end{subequations}
hold for all $(\dw^\mathrm{S}, \dq^\mathrm{S})$, where both
$(\bld{u}^\mathrm{S}, p^\mathrm{S})$ and $(\dw^\mathrm{S}, \dq^\mathrm{S})$
satisfy~\eqref{eq:subscale_def}. This problem implicitly defines the sub-scale
velocity $\bld{u}^\mathrm{S}$ as a function of the macro-scale pressure
$\bar{\bld{g}}$ and the body force $\rho\bld{b}$, postulated formally
in~\eqref{materialLaw}.

Finally let us remark that~\eqref{subscaleProblem1} does not contain the
boundary contributions 
\begin{align}
\int_{\partial \omgb}
\dw^\mathrm{S}
\cdot
\bld{T}, \nonumber
\end{align}
where $\bld{T}$, in analogy to~\eqref{eq:traction_def}, denotes the tractions at
the RVE boundary associated with the stress $\bld{\tau}$ and pressure $p$
from~\eqref{pressureExpansion}. As shown
in~\cite[Eq.~(32)]{SandstromLarsson}, this term vanishes as a consequence
of the periodicity enforced in~\eqref{eq:subscale_def}.

\subsection{Stokes-Darcy coupling}\label{coupling}

\begin{figure}
\centering\includegraphics{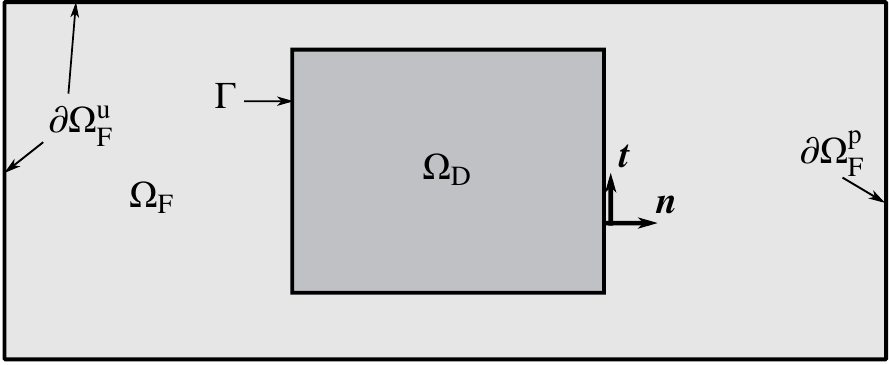}
\caption{\textit{Description of the Stokes-Darcy problem}}
\label{Stokes_Darcy_coupling_pic}  
\end{figure}

A closer inspection reveals that the only terms in~\eqref{eq:weak_stokes}
remaining for further considerations are 
\begin{align}\label{eq:Stokes-Darcy}
\int_\Gamma\left.
\dw
\cdot 
\bld{\tau}\right|_{\Omega_\mathrm{F}}
\cdot\bld{n}\ds
-
\int_\Gamma\left.
\dw
\cdot 
p \right|_{\Omega_\mathrm{F}}
\cdot\bld{n}\ds;
\end{align}
recall Eq.~\eqref{weak_stokes3} in particular. This contribution is made
explicit by imposing the interface conditions between Darcy and Stokes
sub-domains in the form
\begin{subequations}\label{eq:BJS_conditions}
\begin{align}
u_n - \bar{u}_n & = 0  
& \text{on } \Gamma,
\label{BJS_conditions1} \\
p - \bar{p} 
& = 
\bld{n} \cdot \bld{\tau} \cdot \bld{n} 
& 
\text{on } \Gamma,\label{BJS_conditions2}
\\
\beta \left( u_t - \bar{u}_t \right) 
& = \bld{t} \cdot
\bld{\tau} \cdot \bld{n} 
& 
\text{on } \Gamma,
\label{BJS_conditions3}
\end{align}  
\end{subequations}
where $\bld{n}$ and $\bld{t}$ denote the unit vectors normal to and tangent to
$\Gamma$, respectively, see Fig.~\ref{Stokes_Darcy_coupling_pic}, and where we
have utilized the decompositions of the fluid, $\bld{u}$, and seepage, $\bar{\bld{u}}$, velocities into the
components normal and tangential to $\Gamma$,
\begin{align}\label{eq:velocity_decomposition}
\bld{u} 
= 
u_n \bld{n} 
+ 
u_t \bld{t}, &&
\bar{\bld{u}} = \bar{u}_n \bld{n} + \bar{u}_t \bld{t} 
&&
\text{on } \Gamma;
\end{align}
notice that in~\eqref{BJS_conditions2} and~\eqref{BJS_conditions3} we omitted
the symbol $\bigl|_{\Omega_\mathrm{F}}$ to keep the notation simple.
Eq.~\eqref{BJS_conditions2} thus represents the continuity of velocity in the
normal direction, \eqref{BJS_conditions1} enforces the equilibrium condition in
the normal direction, and \eqref{BJS_conditions3} is the
Beavers-Joseph-Saffman~(BJS) slip law~\cite{BeaversSaffman,Saffman} involving
the interface constant~$\beta$. As was shown by Saffman~\cite{Saffman}, the
tangential velocities satisfy $\bar{u}_t \ll u_t$, so that we set
\begin{align}\label{eq:vanishing_velocity_t}
\overline{u}_t = 0  \text{ on } \Gamma.
\end{align}

Utilizing the decomposition~\eqref{eq:velocity_decomposition} also for the test
functions $\dw$, the terms in~\eqref{eq:Stokes-Darcy} expand into
\begin{align}
\int_\Gamma
\delta w_t \bld{t}
\cdot
\bld{\tau}\cdot\bld{n}
\ds +
\int_\Gamma
\delta w_n \bld{n}
\cdot
\bld{\tau}\cdot\bld{n}\ds 
-
\int_\Gamma
\delta w_n p \ds & = 
\nonumber\\
\int_\Gamma
\delta w_t 
\beta 
\left( u_t - \bar{u}_t \right) \ds 
+
\int_\Gamma
\delta w_n
(p - \bar{p})
\ds 
-
\int_\Gamma
\delta w_n p \ds 
& = 
\nonumber\\ 
\int_\Gamma 
\delta w_t
\beta 
u_t \ds 
- 
\int_\Gamma
\delta w_n 
\bar{p}\ds. \label{substitutionOfBJS}
\end{align}

\begin{figure}[ht]
\centering\includegraphics{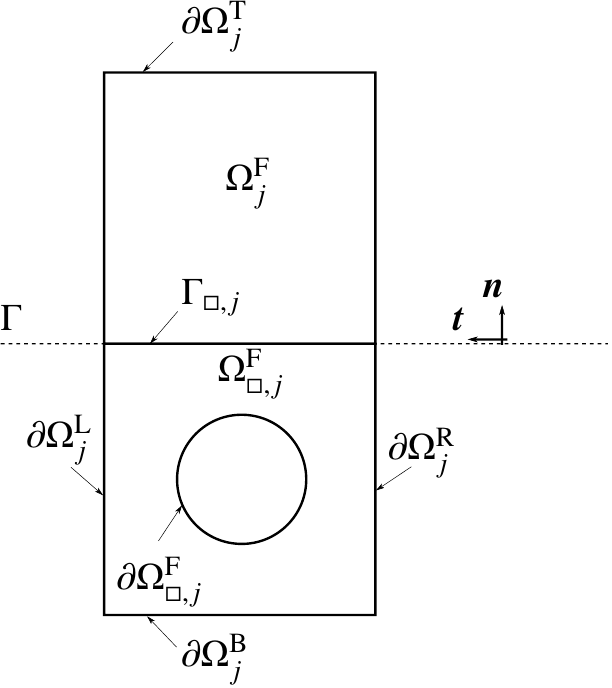}
\caption{\it Scheme of the boundary layer problem.}
\label{fig:blp}
\end{figure}

As follows from analytical results by J\"{a}ger and
Mikeli\'{c}~\cite{JagerMikelic2000} and numerical studies by J\"{a}ger et
al.~\cite{Jager:2001:AAL} and Carraro et al.~\cite{CarraroMikelic2013},
parameter $\beta$ governing the friction in~\eqref{BJS_conditions2} for
Newtonian fluids can be expressed as
\begin{align}\label{computingbeta}
\beta = -\frac{\mu}{C^\mathrm{bl}},
\end{align}
where $\mu$ is viscosity of the fluid, cf.~\ref{eq:Newton}, and
$C^\mathrm{bl}$ is determined from the solution of a boundary layer
problem defined for RVEs in the vicinity of the internal boundary $\Gamma$, as
shown in Fig.~\ref{fig:blp}.\footnote{To be more specific, in the original
work~\cite{JagerMikelic2000}, the boundary layer problem is defined on an
infinite vertical row of RVEs. However, the numerical results with guaranteed
error presented in~\cite{Jager:2001:AAL,CarraroMikelic2013} reveal that the
analysis on the truncated domain in Fig.~\ref{fig:blp} provides sufficiently accurate
results.}

In our notation, the problem reads as: Find $(\bld{u}^\mathrm{S},p^\mathrm{S})$
such that
\begin{align}
  &\int_{\Omega^\mathrm{F}_{\Box,j} \cup \Omega^\mathrm{F}_j}
  \bld{\nabla} \dw^\mathrm{S} : \bld{\nabla} \du^\mathrm{S}
  \dx
  -
  \int_{\Omega^\mathrm{F}_{\Box,j} \cup \Omega^\mathrm{F}_j}(
  \bld{\nabla}\cdot\dw^\mathrm{S})p^\mathrm{S} \dx\nonumber
  \\
+ & 
  \int_{\Omega^\mathrm{F}_{\Box,j} \cup \Omega^\mathrm{F}_j}
  \dq^\mathrm{S} (\bld{\nabla} \cdot \dw^\mathrm{S}) \dx = 0\nonumber
\end{align}
holds for all $(\dw^\mathrm{S},\dq^\mathrm{S})$. All fields above are periodic
at $\partial \Omega_j^\mathrm{L}$ and $\partial \Omega_j^\mathrm{R}$, and satify
in addition $\dw^\mathrm{S} = \du^\mathrm{S} = \bld{0}$ on $\partial
\Omega_{\Box,j}^\mathrm{F}$ and $\partial \Omega_j^\mathrm{B}$,
$\delta w_t^\mathrm{S} = u_t^\mathrm{S} = 0$ on $\partial \Omega_j^\mathrm{T}$,
$\dw^\mathrm{S} = \du^\mathrm{S} = \bld{0}$ on $\Gamma_{\Box,j}$, 
$\llbracket \dw^\mathrm{S} \rrbracket = \llbracket \du^\mathrm{S} \rrbracket =
\bld{0}$ on $\Gamma_{\Box,j}$, and  
\begin{align*}
\int_{\Gamma_{\Box,j}}
\dq^\mathrm{S}
\ds
=
\int_{\Gamma_{\Box,j}}
p^\mathrm{S} \ds
=
0.
\end{align*}
The interface constant $C^\mathrm{bl}$ is then obtained as
\begin{align}
C^\mathrm{bl} 
= 
\int_{\Gamma_{\Box, j}}
u^\mathrm{S}_t \ds,\label{C_bl}
\end{align}
and therefore depends only on the geometry of perforations. To the best of our
knowledge, no extension of these results to non-linear fluids is currently
available.

\subsection{Darcy-Stokes system}

Collecting the results of the previous
Sections~\ref{sec:vartional}--\ref{coupling}, we recast the original
weak formulation~\eqref{eq:weak_stokes} into the final form: Find a quadruple
$(\bld{u},\bar{\bld{u}}, p,\bar{p})$ such that
\begin{align}
& 
\int_{\omgd} 
\bdq 
(\bld{\nabla} 
\cdot 
\bar{\bld{u}})
\dx 
+ 
\int_{\omgd}
\dwD\cdot\bar{\bld{u}}
\dx 
- 
\int_{\omgd}
\dwD \cdot 
\bld{\bar{w}}(\bld{\nabla} \bar{p}, \rho \bld{b})\dx
\nonumber
\\ 
+ & 
\int_{\omgf}
\bld{\nabla} \dw : \bld{\tau}(\bld{D}(\bld{u})) 
\dx 
- 
\int_{\omgf}( \bld{\nabla}\cdot\dw) p \dx
+ 
\int_{\omgf} 
\dq (\bld{\nabla} \cdot \bld{u}) 
\dx\nonumber 
\\ 
+ & \int_{\pomgpf} \delta w_n\hat{p} \ds
+
\int_\Gamma 
\delta w_t \beta u_t 
\ds 
- 
\int_\Gamma
\delta w_n 
\bar{p}
\ds\nonumber \\ 
= & \int_{\omgf}\dw \cdot \rho \bld{b} \dx
\label{st_da_final} 
\end{align}
for all $(\dw,\dwD,\dq,\delta \overline{q})$, where $(\bld{u},\dw) =
(\hat{u}_n \bld{n}, \bld{0})$ on $\pomgvf$ according to~\eqref{BC_gmv},
$\bar{u}_t = \delta\bar{w}_t = 0$ on $\Gamma$ according
to~\eqref{eq:vanishing_velocity_t}, and $\bar{u}_n = u_n$ and $\delta \bar{w}_n
= \delta w_n$ on $\Gamma$ by~\eqref{BJS_conditions1}, cf.
Fig.~\ref{Stokes_Darcy_coupling_pic}.

\section{Numerical solution}\label{sec:numerical_solution}

Notice that the problem~\eqref{st_da_final} is non-linear, because of possible
non-linearity of the constitutive laws $\bld{\tau}(\bld{D}(\bld{u}))$ and
consequently of $\bld{\bar{w}}(\bld{\nabla} \bar{p}, \rho \bld{b})$, and as such
it is treated using successive consistent linearization introduced in
Section~\ref{sec:linearization}. The finite element discretization of the
ensuing linear system and the implementation of a Newton solver are outlined in
Section~\ref{sec:finite_elements}.

\subsection{Linearization}\label{sec:linearization}

In order to avoid a profusion of notation, we denote still by
$(\bld{u},\bar{\bld{u}}, p,\bar{p})$ the solution around which the linearization
is performed. Its iterative correction $(\bld{v},\bar{\bld{v}}, q, \bar{q})$
is defined by the associated linear problem,
cf.~\eqref{st_da_final}: 
\begin{align}
&
\underbrace{ \int_{\omgd} \bdq (\bld{\nabla}\cdot \bbld{v}) \dx}_{\bar{G}} 
+ 
\underbrace{\int_{\omgd}\bld{\dwD} \cdot \bbld{v} \dx}_{\bar{M}}
-
\underbrace{\int_{\omgd}\bld{\dwD} 
\cdot
\frac{
\partial\bld{\bar{w}}(\bld{\nabla}
\bar{p},
\rho\bld{b})}{\partial\bld{\nabla}\bar{p}}\cdot\bld{\nabla}\bar{q}
\dx}_{\bar{K}_t}\nonumber\\ + & \underbrace{\int_{\omgf} \bld{\nabla} \dw
:\frac{\partial \bld{\tau}(\bld{D})}{\partial
\bld{\nabla}\bld{u}}:\bld{\nabla}\bld{v}\dx}_{K_t} -
\underbrace{\int_{\omgf}(\bld{\nabla} \cdot \dw)q \dx}_{G_p} +
\underbrace{\int_{\omgf}\dq(\bld{\nabla} \cdot \bld{v}) \dx}_{G_u}\nonumber \\
&  + \underbrace{\int_{\Gamma} \delta w_t\beta v_t
\ds}_{M_\Gamma}- \underbrace{\int_{\Gamma}\delta w_n\bar{q}
\ds}_{G_\Gamma}\nonumber \\         = &  \label{LinearizedStokesDarcy} \\ - &  
\underbrace{\int_{\omgd} \bdq (\bld{\nabla}\cdot \bbld{u}) \dx}_{\bar{f}_q}
-\underbrace{\int_{\omgd}\bld{\dwD} \cdot \bbld{u} \dx}_{\bar{f}_u} + \underbrace{\int_{\omgd}\bld{\dwD} \cdot \bld{\bar{w}}(\bld{\nabla}\bar{p}, \rho \bld{b}) \dx}_{\bar{f}_w}\nonumber\\ & -\underbrace{\int_{\omgf} \bld{\nabla} \dw : \bld{\tau}(\bld{D}) \dx}_{f_\tau}+ \underbrace{\int_{\omgf}(\bld{\nabla} \cdot \dw)p \dx}_{f_p} - \underbrace{\int_{\omgf}\dq(\bld{\nabla} \cdot \bld{u}) \dx}_{f_q} \nonumber \\ & - \underbrace{\int_{\Gamma} \delta w_t\beta u_t \ds}_{f_f} + \underbrace{ \int_{\Gamma}\delta w_n\bar{p} \ds}_{f_\Gamma} + \underbrace{\int_{\omgf}\dw \cdot \rho \bld{b} \dx}_{f_b}+ \underbrace{\int_{\pomgpf}\delta w_n \hat{p} \ds}_{f_t},\nonumber
\end{align}
where $\bld{v} = \bld{0}$ on $\pomgvf$, and the seepage
velocity $\bar{\bld{v}}$ and the test functions $(\dw,
\bld{\delta}\bar{\bld{w}}, \dq, \bdq)$ satisfy the same conditions as in the
non-linear problem in~\eqref{st_da_final}. 

The linearization is finalized by evaluating the sensitivities of the local
constitutive laws. In the Stokes domain, the application of the chain rule
gives
\begin{align}\label{eq:Stokes_diff}
\frac{\partial\bld{\tau}(\bld{D})}{\partial(\bld{\nabla}\bld{u})} 
=  
\frac{\partial\bld{\tau}(\bld{D})}{\partial\bld{D}}:\frac{\partial\bld{D}}{\partial(\bld{\nabla}\bld{u})} 
= 
\frac{\partial\bld{\tau}(\bld{D})}{\partial\bld{D}}:\bld{\curly{I}_{dev}},
\end{align}
where the specific form of the derivative
$\frac{\partial\tau(\bld{D})}{\partial\bld{D}}$ depends on the particular choice
of constitutive law and $\bld{\curly{I}_{dev}}$ is the deviatoric projector defined as
\begin{align}
[\bld{\curly{I}_{dev}}]_{ijkl} = \delta_{ik}\delta_{jl} - \frac{1}{3}\delta_{ij}\delta_{kl}.
\end{align}
In the homogenized Darcy
domain, it holds~\cite[Section~5]{SandstromLarsson}
\begin{align}
\frac{\partial\bld{\bar{w}}(\bld{\nabla}
\bar{p},
\rho\bld{b})}{\partial \bld{\nabla}
\bar{p}}
=
\phi \sum_{k=1}^{2} \left < \bld{u}^\mathrm{S}_{(k)} \right > \otimes
\bld{e}_{(k)},\label{permeabilityMatrix}
\end{align}
where $\bld{u}^\mathrm{S}_{(k)}$ denotes the solution to the linearized
sub-scale problem~\eqref{subscaleProblem} with the macroscopic pressure gradient
$\bar{\bld{g}}$ set to the $k$-th canonical basis vector $\bld{e}_{(k)}$.

\subsection{Finite element treatment}\label{sec:finite_elements}

The linearized weak form can now be discretized using
conventional finite element procedures. At the element level,
Eq.~\eqref{LinearizedStokesDarcy} is recast into the equivalent matrix form
\begin{equation}\label{discretized_Darcy}
\begin{pmatrix}
\bld{K}_t + \bld{M}_\Gamma + \bar{\bld{M}} &\bar{\bld{K}}_t + \bld{G}_\Gamma + \bld{G}_p\\
\bld{G}_u + \bar{\bld{G}} & \bld{0}_{3 \times 3}
\end{pmatrix}
\begin{Bmatrix}
\bld{v} \\
\bld{q}
\end{Bmatrix}
=
\begin{Bmatrix}
\bld{f}_{ext,u} -\bld{f}_{int,u} \\
-\bld{f}_{int,p}
\end{Bmatrix},
\end{equation}
where the individual terms contain contribution from in the Darcy domain $\omgd$, the Stokes
domain $\omgf$, at the internal interface $\Gamma$, or at the boundary $\gmp$,
in accordance with~\eqref{LinearizedStokesDarcy}. The vector on the right hand side consists of sub-vectors defined as
\begin{align}
&\bld{f}_{ext,u} = \bld{f}_t + \bld{f}_b,\nonumber \\
&\bld{f}_{int,u} = \bar{\bld{f}}_q+\bar{\bld{f}}_u-\bar{\bld{f}}_w+\bld{f}_\tau-\bld{f}_p+\bld{f}_f-\bld{f}_\Gamma,\qquad \bld{f}_{int,p} = \bld{f}_q\nonumber
\end{align}

Notice the $3 \times 3$ null
sub-matrix in~\eqref{discretized_Darcy} which may cause numerical problems, if
spaces of approximation functions do not satisfy the so-called inf-sup
condition~\cite{LBB, LBB_Babuska}.
For this reason, we employed the Taylor-Hood $P2/P1$ elements with the quadratic
approximation of velocity and the linear approximation of pressure. Depending on the order of the polynomial in individual terms, 3 or 7 integration points were used. The sub-scale problem~\eqref{subscaleProblem} is treated in the same fashion. The permeability matrix (\ref{permeabilityMatrix}) is being kept constant for each one element inside the Darcy Domain, i.e. the sub-scale problem is solved only once (in one integration point) for each Darcy element.

At the structural level, we use the Newton method with a back-tracking
line search to ensure a robust convergence with the quadratic rate. In
particular, we used the backtracking algorithm introduced
in~\cite[Section~7.2.3, Program~62]{Quarteroni}, with the decrease parameter
$\sigma = 10^{-4}$ and with the scaling factor $\rho = 0.5$. The outlined
solution procedure was implemented into OOFEM code~\cite{PatzakBittnarOOFEM,PatzakOOFEMActa}.

\section{Examples}\label{NumericalExamplesSec}

In this section, two benchmark tests are presented to illustrate the
capabilities and performance of the proposed method. The first example,
Fig.~\ref{uniflow_Scheme}, illustrates unidirectional flow of a fluid around reinforcing bars. The second example,
Fig.~\ref{second_test_Scheme}, is more complex and illustrates the complex flow
of the concrete over the reinforced area with a significant effect of the
friction interface parameter $\beta$, recall Eq.~\eqref{BJS_conditions3}. In
both examples, the solution based on the homogenization technique is verified
against a fully resolved solution computed by Direct Numerical Simulation~(DNS).

We will  consider two constitutive laws: Newtonian fluids with
viscosity $\mu$, 
\begin{equation}\label{eq:Newton}
\bld{\tau} = \mu \bld{D},
\end{equation}
and a regularized Bingham model~\cite{Papanastasiou} defined as 
\begin{align}
    \bld{\tau} = \left [ \mu_0 + \frac{\tau_0}{\J}\left (1 - \exp(-m\J) \right)
    \right ] \bld{D} \label{binghamfluid},
\end{align}
where $\tau_0$ is the yield stress, representing an initial resistance to the
flow, $\mu_0$ is the plastic viscosity, which governs the flow once the
yield stress threshold has been passed, and $m$ is the regularization parameter.
The second invariant of the deviatoric strain tensor $J_{2}^{D}$ is defined as 
\begin{align}
J_{2}^{D}=\frac{1}{2}\boldsymbol{D}:\boldsymbol{D}, \nonumber
\end{align}
The constitutive tangents, needed in~\eqref{eq:Stokes_diff}, are thus provided
by 
\begin{align}
\frac{\partial\bld{\tau}(\bld{D})}{\partial\bld{D}} & =
\mu_\mathrm{app}(\bld{D}) \bld{\curly{I}}_{\mathrm{sym}} +
\mu'_\mathrm{app}(\bld{D})
\bld{D}\otimes\bld{D}\label{BinghamDerivative},
\end{align}
where $\bld{\curly{I}}_\mathrm{sym}$ is the unit fourth order tensor with both
major and minor symmetries, and the apparent viscosity $\mu_\mathrm{app}$ and
its derivative $\mu'_\mathrm{app}$ are provided by
\begin{subequations}
 \begin{align}
\mu_\mathrm{app}(\bld{D}) & = \mu_0 + \frac{\tau_0}{\J}\left (1 - \exp(-m\J)
\right), \nonumber\\ 
\mu'_\mathrm{app}(\bld{D}) & = \frac{\left[ \left ( m\J + 1 \right)\exp\left(-m\J\right) - 1\right ]\tau_0}{2(\JJ)^{\frac{3}{2}}}.\nonumber
\end{align}
\end{subequations}
In what follows, all quantities will be expressed in consistent units. 

We also wish to emphasize that both examples involve an array of only $4 \times
4$ RVEs of unit length with the obstacle radius of $\xi$, so that macro and micro
length scales are not widely separated. This setup is, therefore, quite
unfavorable for the multi-scale method, but our results show that it still
corresponds well with DNS. Even higher accuracy is expected for an increasing
length scale contrast.

\subsection{Unidirectional flow}
\begin{figure}[b]
\centerline{\includegraphics{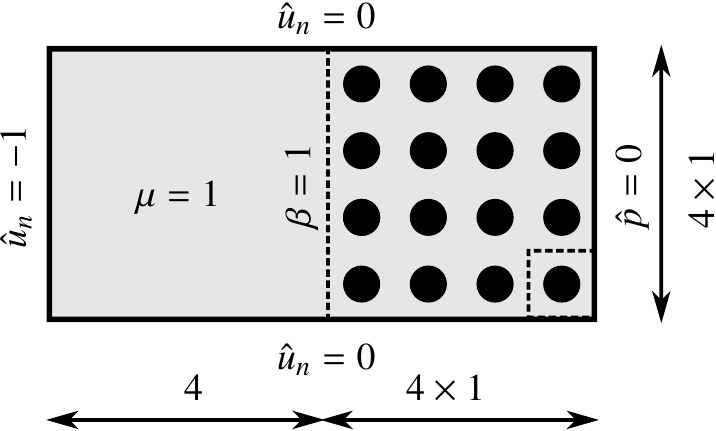}}
\caption{\textit{Unidirectional flow: scheme of the benchmark test.}}
\label{uniflow_Scheme}
\end{figure}

\begin{figure}[h]
\centerline{
 \begin{tabular}{cc}
\includegraphics[width=.425\textwidth]{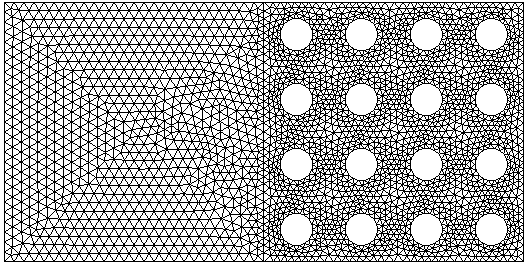}
&
\includegraphics[width=.425\textwidth]{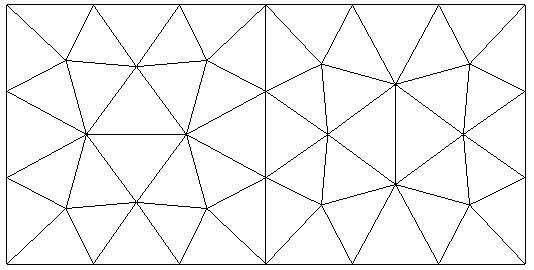}
\end{tabular}
}
\caption{\textit{Unidirectional flow: comparison of the meshes. Mesh for fully
resolved problem (left - $11,358$ nodes, $5,004$ quadratic elements) and for
homogenized problem (right - $130$ nodes, $52$ quadratic elements).}}
\label{uniflow_meshes}
\end{figure}

The setup of this benchmark test is shown in Fig.~\ref{uniflow_Scheme}. The flow
is driven by constant velocity in the normal direction, prescribed on the left
edge of the domain. The top and the bottom edges are friction-free and the ``do
nothing'' boundary condition is prescribed on the right edge. The Newton fluid
of the unit viscosity, $\mu = 1$, is considered in this elementary example and
the radius of bars is taken as $\xi = 0.25$. The friction parameter $\beta$ is
set to $1$ on the Stokes side of the interface, but it does not play any
important role in this case, as the flow is perpendicular to the interface.
Computational meshes for both fully resolved and homogenized cases are shown in
Fig.~\ref{uniflow_meshes}.

\begin{figure}[b]
\centerline{
\begin{tabular}{cc}
\includegraphics[width=.475\textwidth]{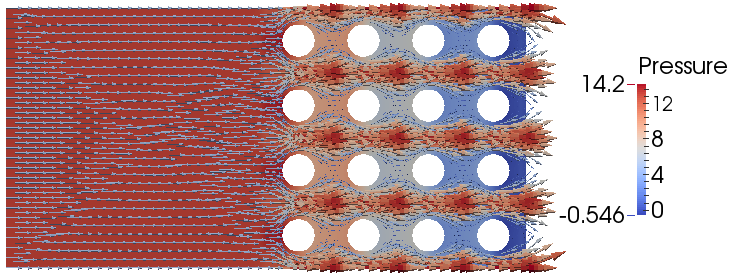}
& 
\includegraphics[width=.475\textwidth]{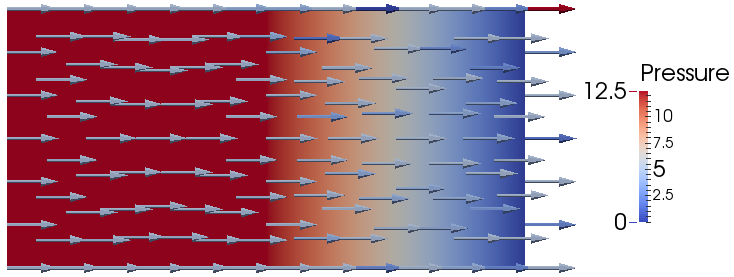}
\end{tabular}
}
\caption{\textit{Unidirectional flow: comparison of fully resolved solution
(left) and homogenized solution (right).}}  
\label{uniflow_Solution}
\end{figure}

In Fig.~\ref{uniflow_Solution}, the results obtained from the homogenized
formulation (left) and the fully resolved solution (right) are presented. The
pressure contours are shown in the background, while the velocity field is
visualized using arrows in nodal points. Notice that the velocity field in the
homogenized solution is uniform and equal to the prescribed boundary condition
due to the incompressibility constraint, as the fluid cannot flow elsewhere but
through the reinforced domain. The pressure distributions are in very good
agreement; the error in the $\max$ norm is approximately~$10\%$.

\subsection{Flow over reinforced area}

\begin{figure}[h]
\centering
\begin{tabular}{cc}
\includegraphics{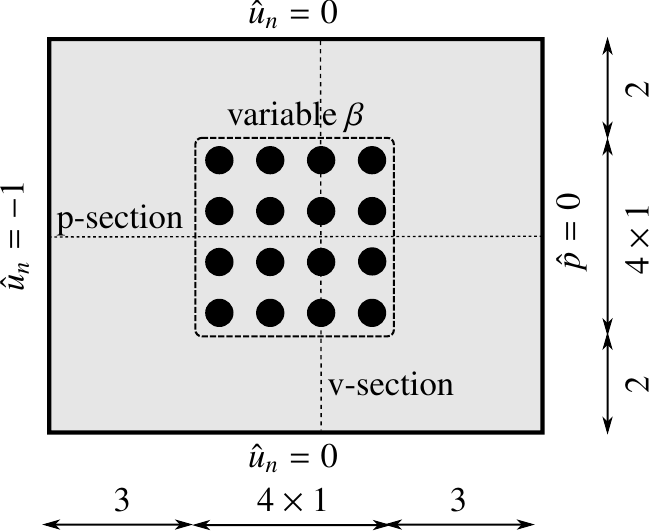}
&
\includegraphics{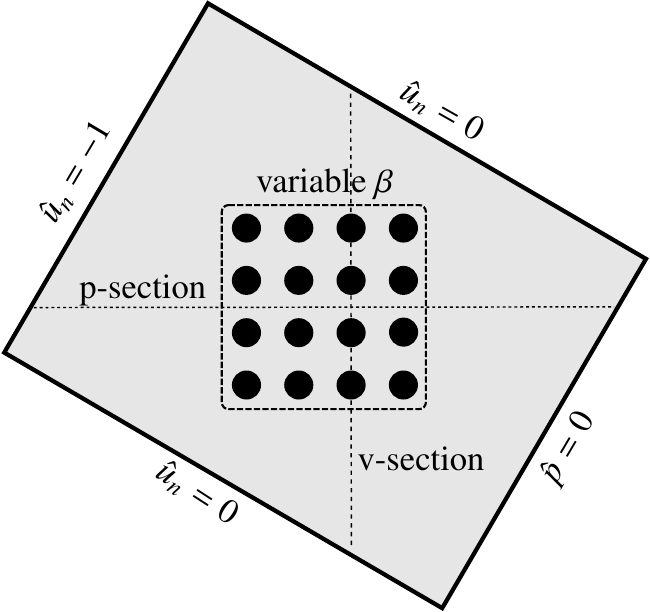}
\end{tabular}
\caption{\textit{Flow over reinforced area: scheme of the benchmark
test.}}
\label{second_test_Scheme}
\end{figure}

The second example illustrates the more complex flow pattern around the
reinforced domain. The reinforced area is located in the middle of the problem
domain, so the fluid is not forced to go through the reinforcing bars and the
whole situation is closer to real casting problems. This problem also allows us
to study the influence of the friction parameter $\beta$. The test is considered
in two variations. In the first variant, the geometry of the reinforced area is
parallel to the direction of the prescribed flow, while in the second case, the
geometry of the reinforced area is rotated with respect to the prescribed flow
direction. The schematic setup of both situations is outlined in
Fig.~\ref{second_test_Scheme}. Again, uniform velocity is prescribed on the left
side, no friction on the top and the bottom and zero ``do nothing'' boundary
condition on the right. The RVEs are chosen with $\xi = 0.125$, $\xi = 0.25$,
and $\xi = 0.35$.

\begin{figure}[b]
\centerline{%
\begin{tabular}{ccc}
\includegraphics[width=.3\textwidth]{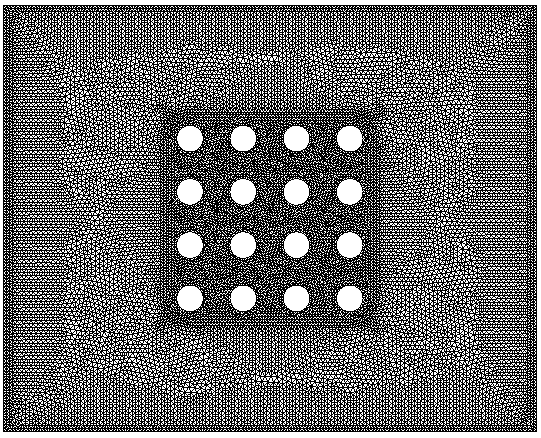} &
\includegraphics[width=.28\textwidth]{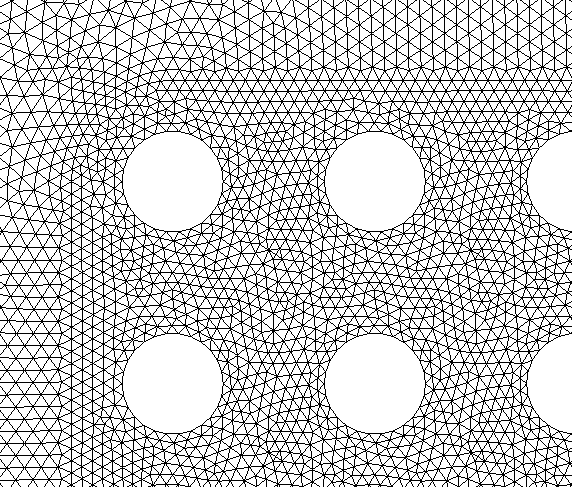} &
\includegraphics[width=.3\textwidth]{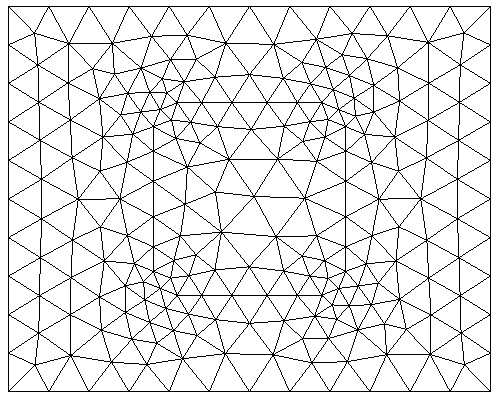}
\end{tabular}
}
\caption{\textit{Flow over reinforced area: comparison of the meshes. The mesh
for DNS with $\xi = 0.25$ is on the left, its detail is in the middle, and the
mesh used for the homogenized approach is on the right.}}\label{meshes_fover}
\end{figure}

\subsubsection{Newtonian flow}

\begin{figure}[h]
\centering
\begin{tabular}{cc}
\includegraphics[width=.425\textwidth]{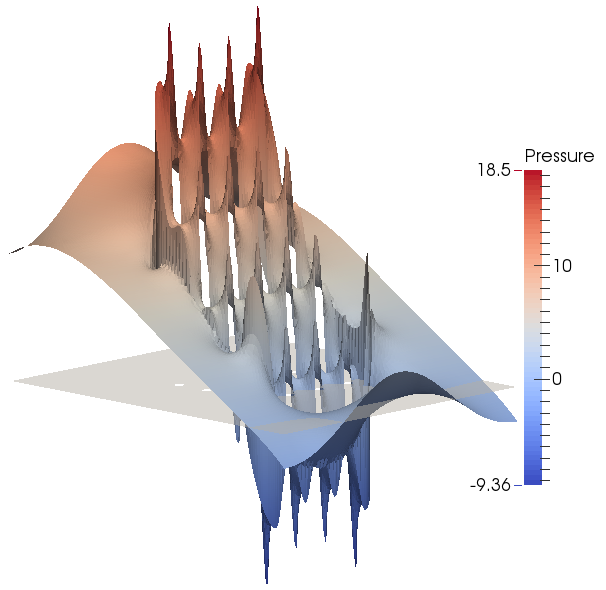} &
\includegraphics[width=.425\textwidth]{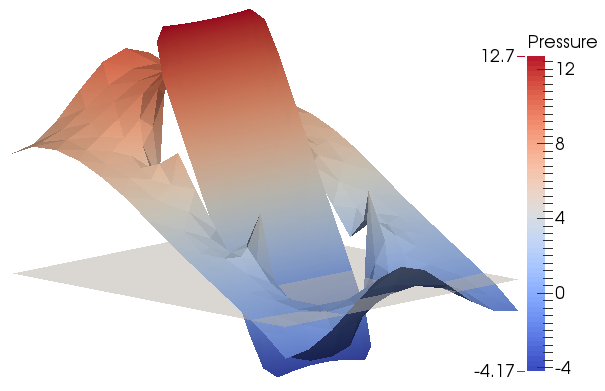} \\
\includegraphics[width=.45\textwidth]{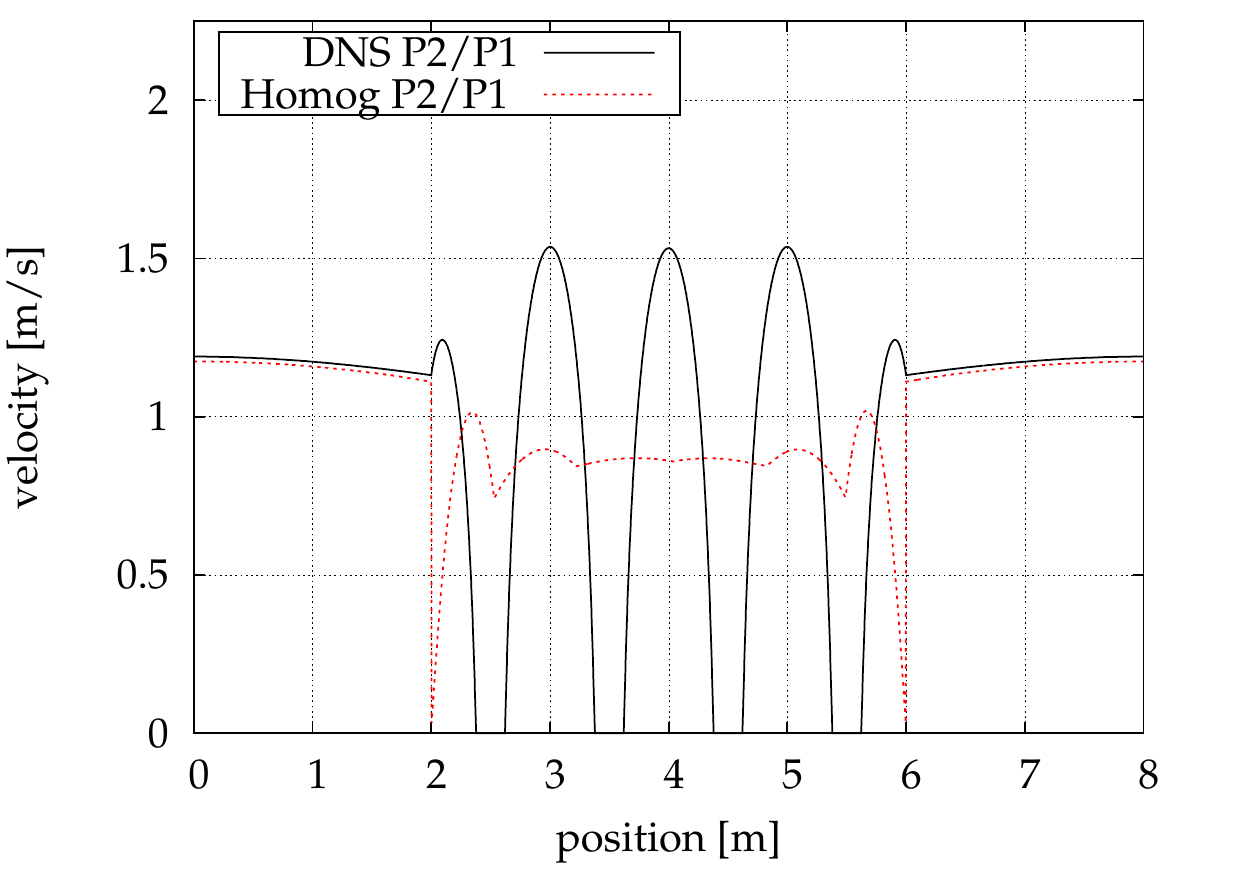} &
\includegraphics[width=.45\textwidth]{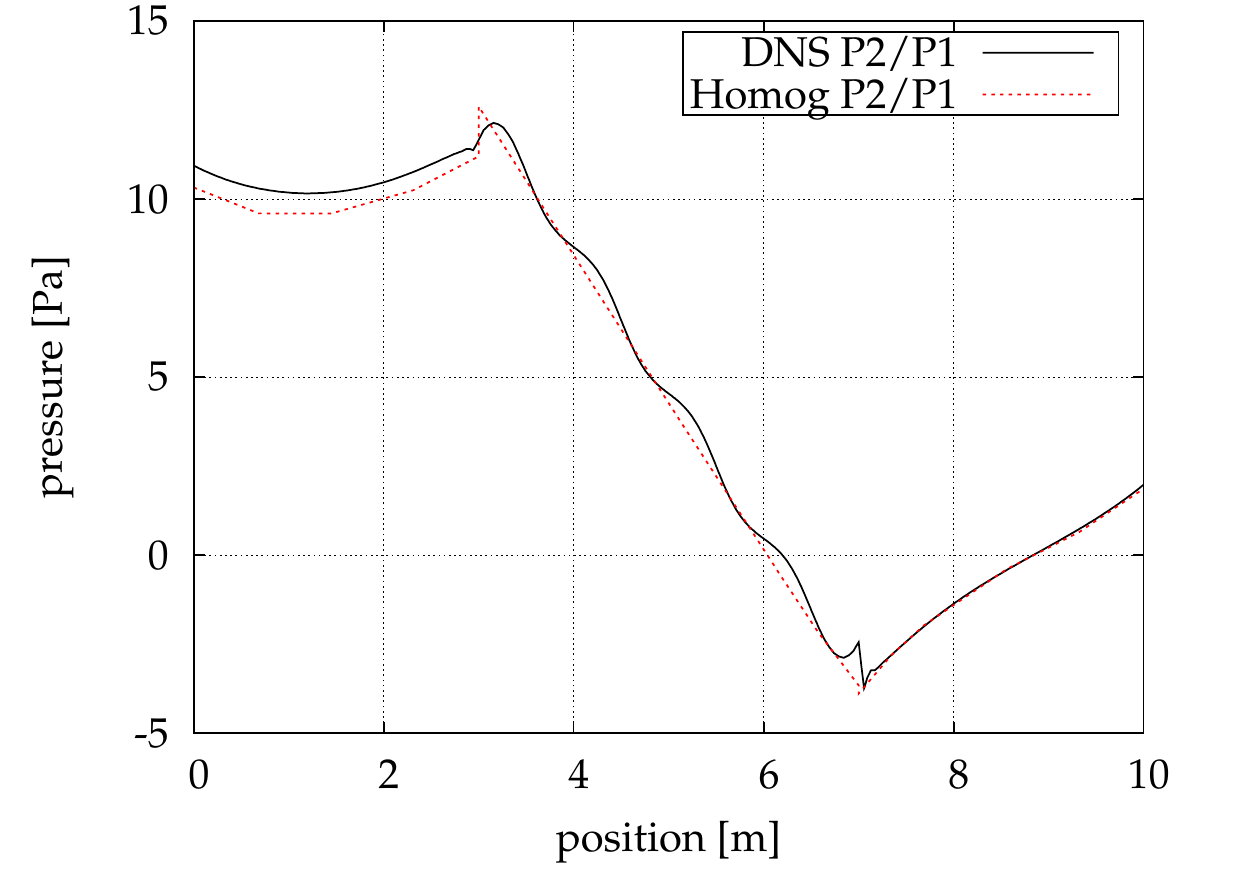}
\end{tabular}
\caption{\textit{Flow over reinforced area: comparison of fully resolved
solution (on the left) and homogenized solution (on the right) for a Newtonian
fluid with $\mu = 20$ and obstacle radius $\xi = 0.125$.
}}\label{second_ql_0.125}
\end{figure}

At first, we concentrate a Newtonian fluid with viscosity $\mu = 20$, the
friction parameter $\beta = 0$, and the obstacle radius $\xi = 0.125$. Results
are collected in Fig.~\ref{second_ql_0.125}. In particular, the first row in
Fig.~\ref{second_ql_0.125} shows an axonometric view of pressure functions. On
the left is the fully resolved pressure function from DNS, on the right is the
pressure function obtained by the homogenization approach. Observe that only the
macroscopic part of the pressure $\bar{p}$ is shown in order to demonstrate that
the pressure gradient is captured extremely well. This can be seen in the second
row of figures, which compares velocity and pressure profiles through the
sections indicated in Fig.~\ref{second_test_Scheme}. The velocity profile is
plotted over the section through the third column of obstacles, and the pressure
profile is plotted over the horizontal section through the whole domain. In both
profiles we compare fully resolved solutions and homogenized solutions. The
error in $\max$ norm in average velocity across the reinforced domain is lower than 1\%. The error in the pressure gradient is less than 5\%.
Fig.~\ref{meshes_fover} illustrates the FE meshes used for homogenized approach
and DNS. In the case of DNS, the mesh was composed of $66,167 $~nodes, while in
the case of the homogenized approach, there were only $961$ nodes, which is
considerably less. The sub-scale problem is solved on $1\times 1$ RVE which is discretized with 745 $P2/P1$ element, which means $1603$ nodes.

\subsubsection{Influence of the friction}

\begin{figure}[p]
\begin{center}
\begin{tabular}{cc}
\includegraphics[width=.48\textwidth]{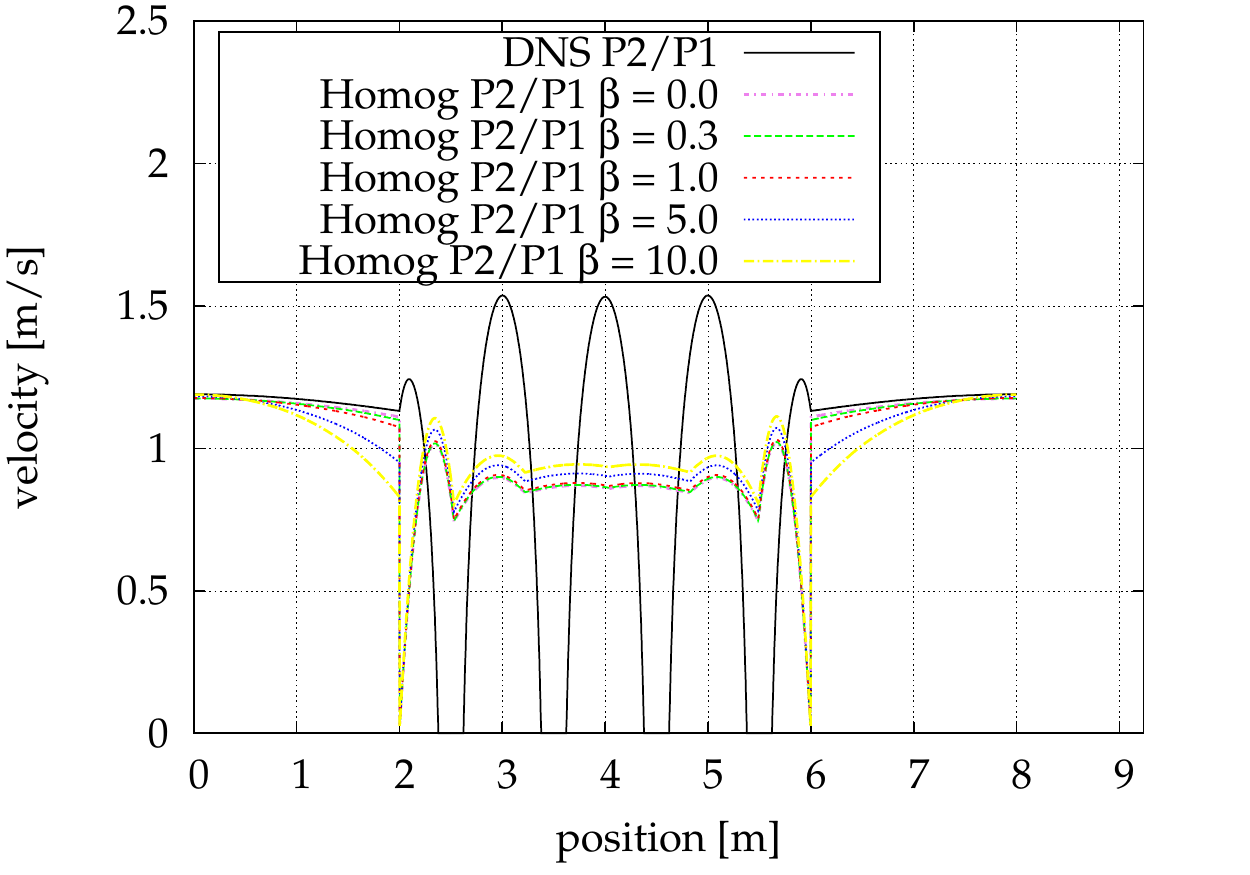} &
\includegraphics[width=.48\textwidth]{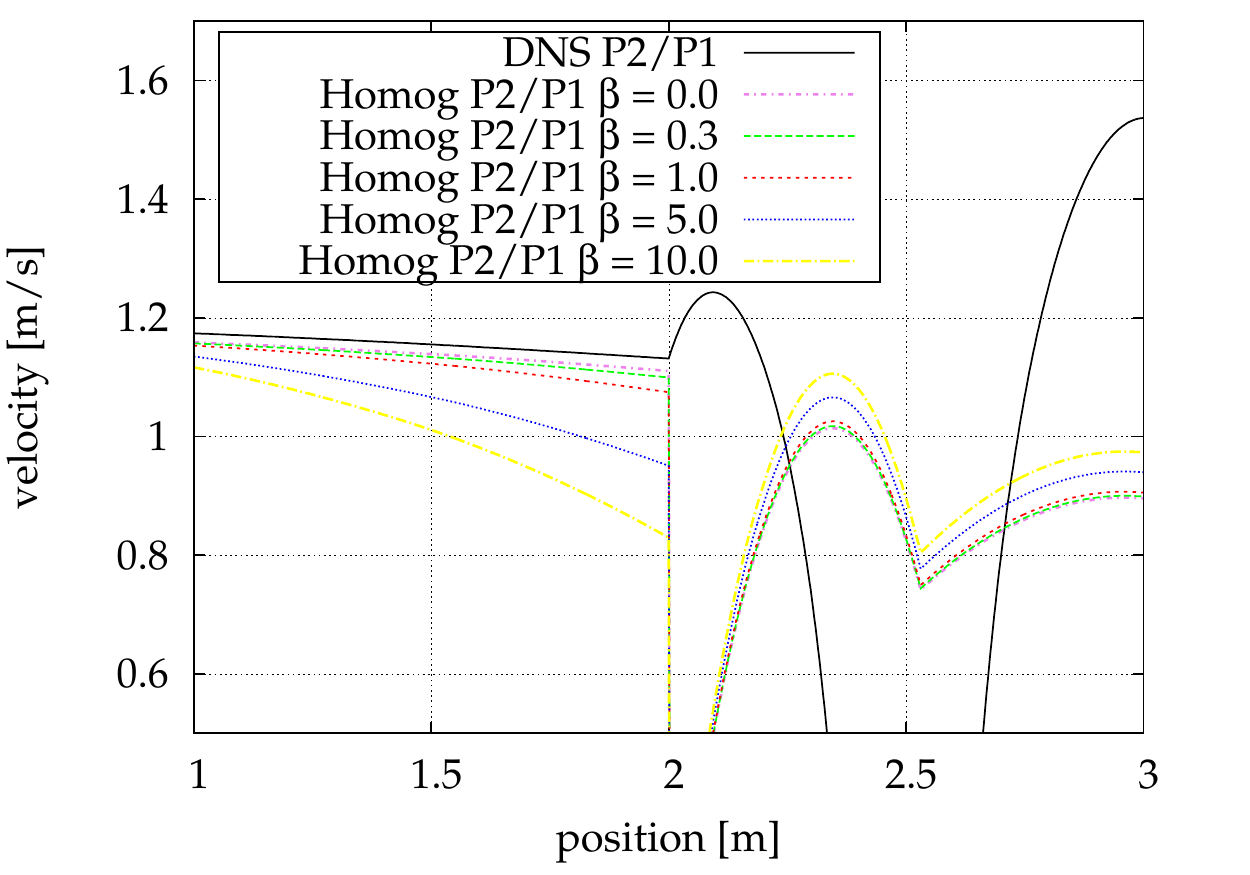} \\ 
\includegraphics[width=.48\textwidth]{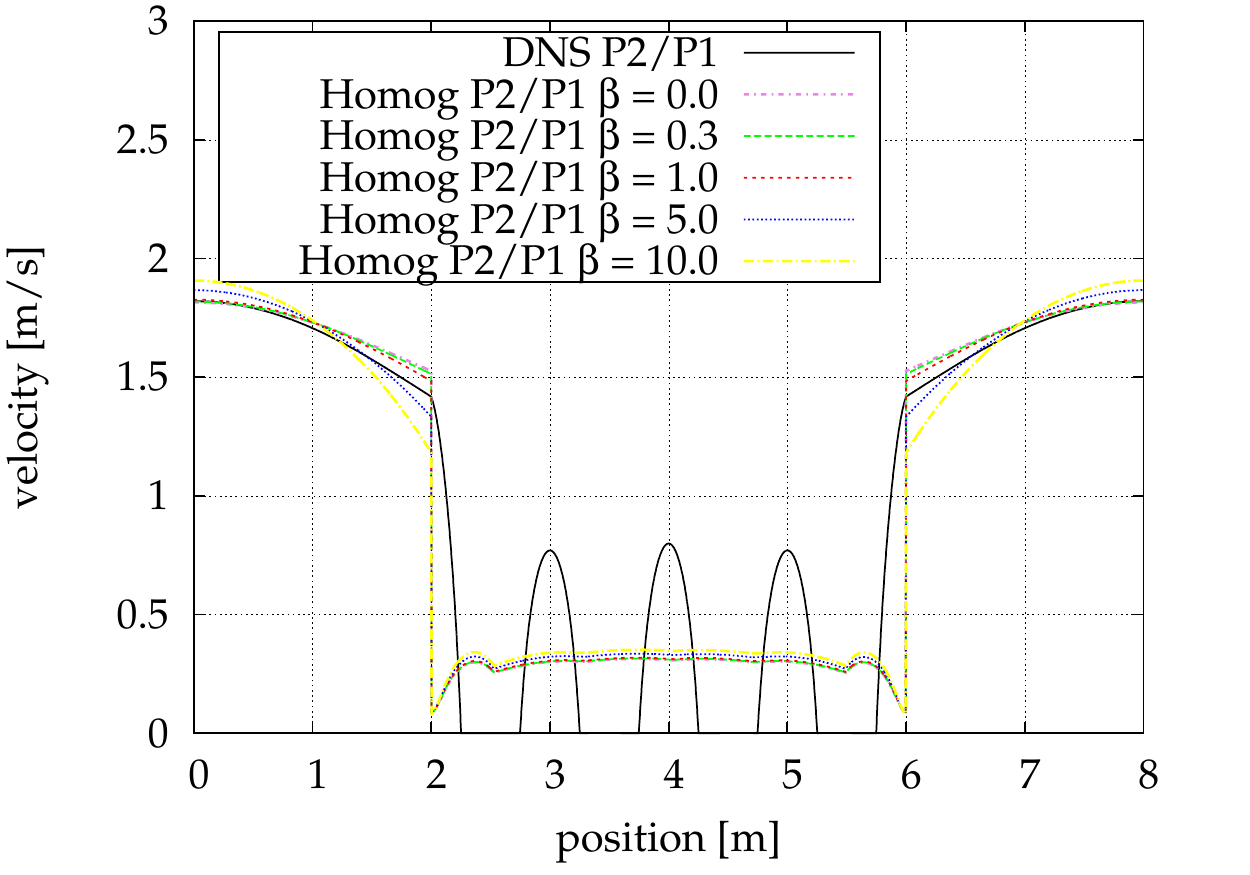} &
\includegraphics[width=.48\textwidth]{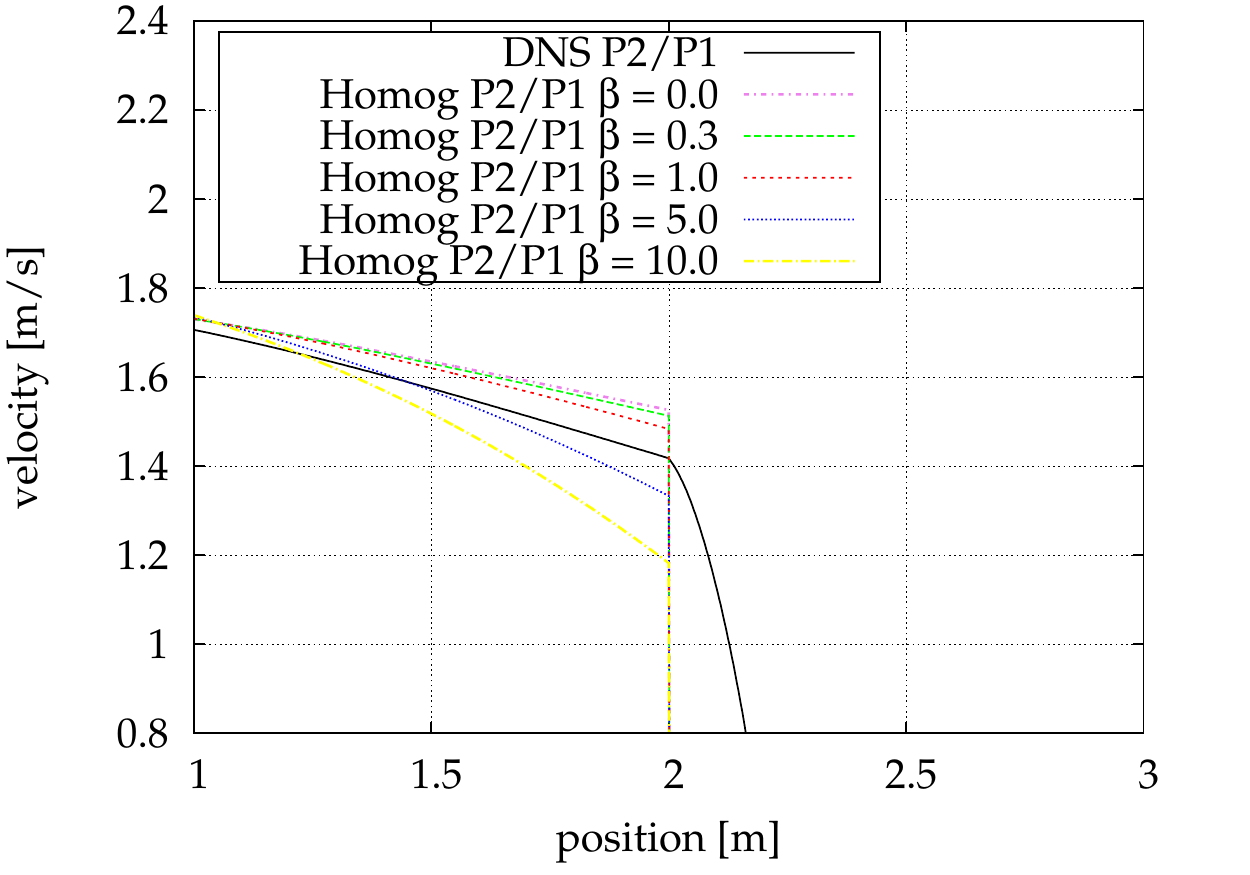} \\ 
\includegraphics[width=.48\textwidth]{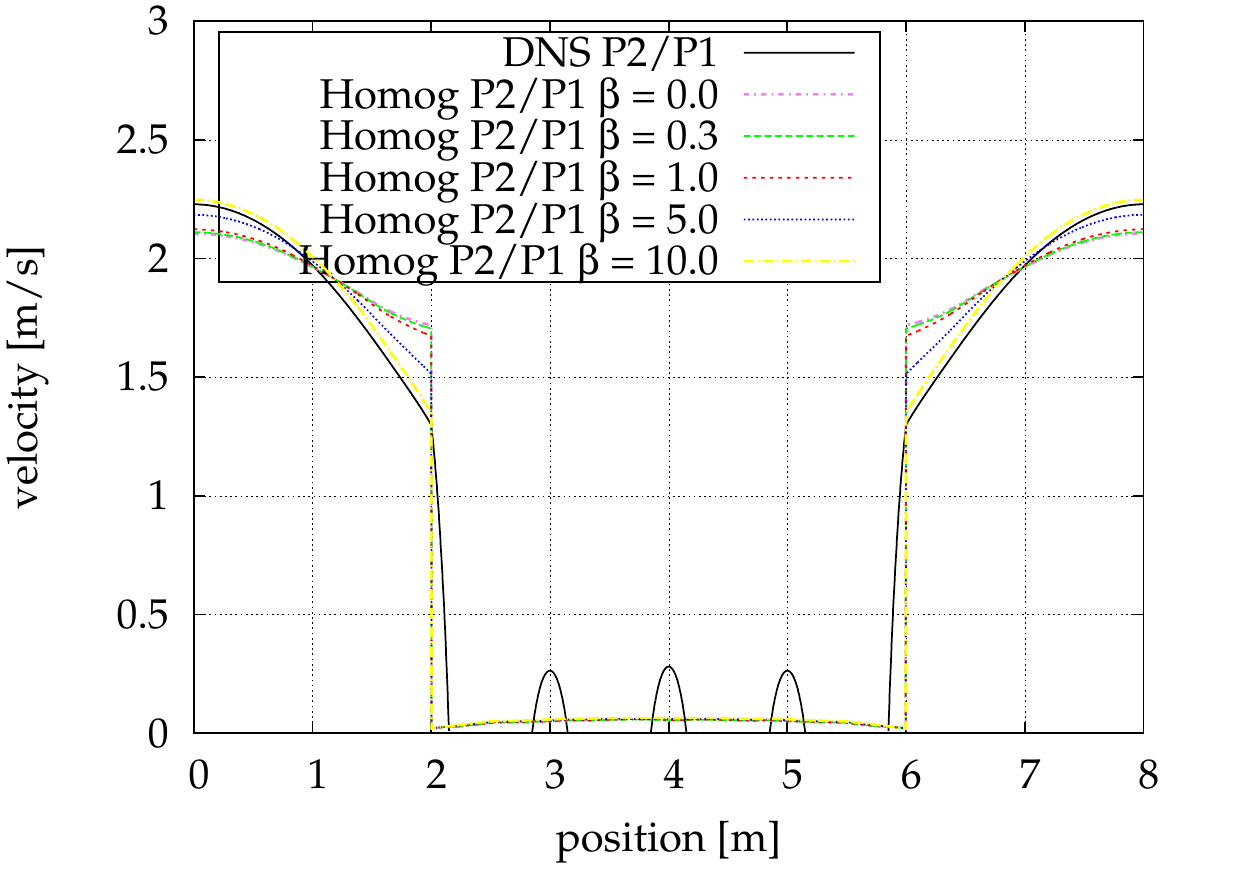} &
\includegraphics[width=.48\textwidth]{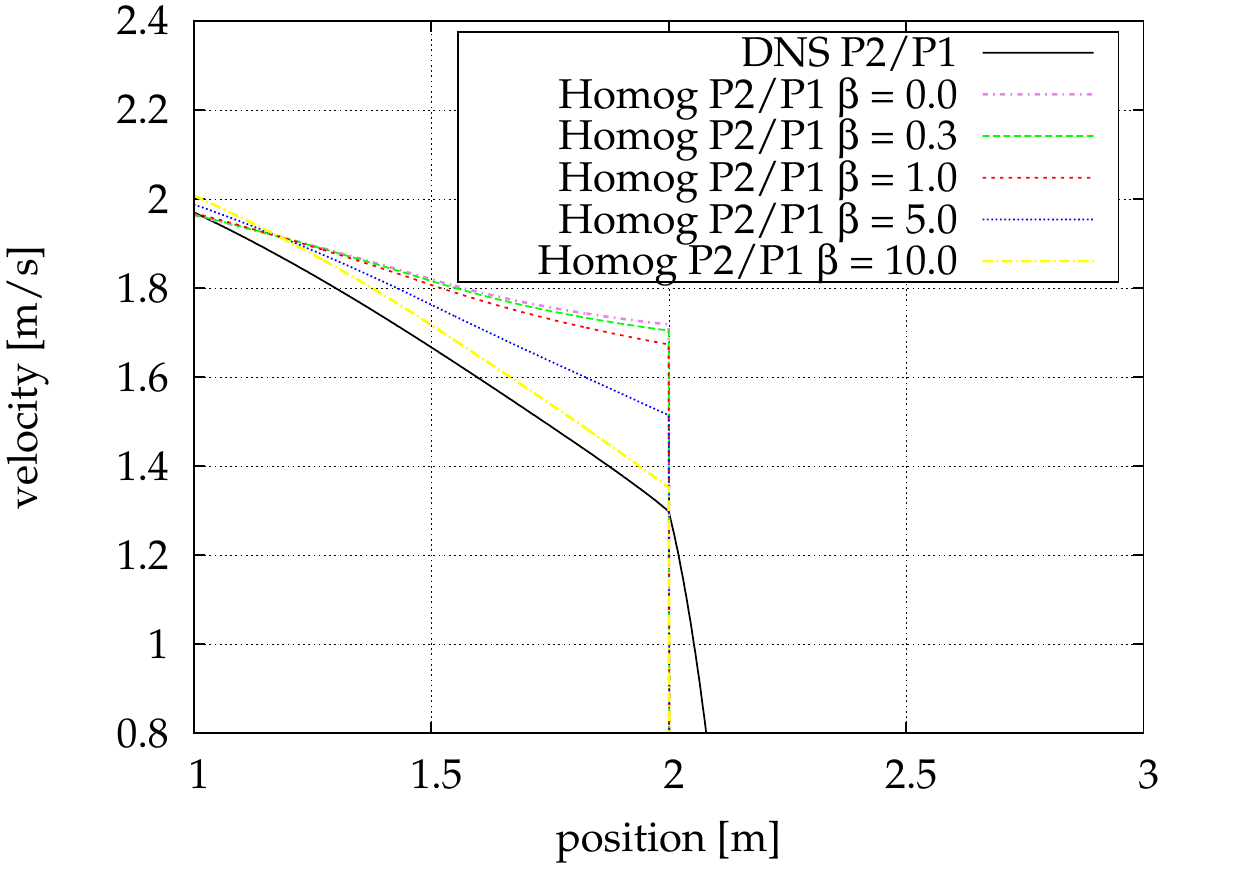} 
\end{tabular}
\caption{\textit{Flow over reinforced area: influence of the friction. The full
velocity profiles are on the left, details near the interface between reinforced
and unreinforced area are on the right. The top line figures show the case of
$\xi = 0.125$, the middle line of $\xi = 0.25$ and the bottom line $\xi =
0.35$.}}\label{friction}
\end{center}
\end{figure}

This subsection is devoted to studying the effect of the friction, for the setup
shown in Fig.~\ref{second_test_Scheme}~(left). The values of parameter $\beta$
are gradually increased from $0$ to $10$, as can be seen from results in
Fig.~\ref{friction}. Velocity profiles in the v-section for different values of
$\beta$ are shown on the left, while detailed view of velocity behavior near the
interface between both the unperforated and the perforated region is shown on
the right. The first row of pictures correspond to $\xi = 0.125$, the second row
to $\xi = 0.25$ and the third to $\xi = 0.35$.

It can be seen that in the case of $\xi = 0.125$, with an increasing value of
$\beta$, the agreement between homogenized and fully resolved solutions
declines. The best agreement appears to be with $\beta = 0$, which corresponds
to no friction at all (full slip conditions). We believe that this phenomenon is
caused by the fact that the effect of prescribed zero velocity on the obstacles
(in the fully resolved case) does not propagate far enough, and so there is no
``friction effect'' at the interface in the homogenized case. Obviously
$\beta=0$ would not be obtained with formula \ref{C_bl}. This is caused by the
particular $4\times4$ setup used in our example, where the obstacles are too far
from the interface $\Gamma$. In the limit case of infinitely many obstacles with
radii tending to zero, significant influence of the friction could be expected.
For a larger size of the obstacles, the situation is different, and with
increasing perimeter of the obstacles, the influence of the friction becomes
more important. For the case of $\xi = 0.25$, the optimal value is $\beta
\approx 3$, and for $\xi = 0.35$, $\beta \approx 10$. Unfortunately, these
values are not in agreement with the results presented
\cite{CarraroMikelic2013}. In our opinion, there are several reasons for that.
First, the geometry of the problem solved by the authors in
\cite{CarraroMikelic2013} is not fully comparable with our problem, as they
consider an infinite periodic domain, where the flow is realized only in the
tangential direction to the interface $\Gamma$. In our case, the perforated
domain has finite dimensions and we also allow the flow through the interface
$\Gamma$.

For that reason and because of the possible material dependency in the case of
non-linear fluids, we rather fit parameter $\beta$ according to agreement in the
velocity profile near the interface. In particular, according to our numerical
experiments~(not shown), the optimal values of $\beta$ determined for a
Newtonian fluid of viscosity $\mu$ performed well also for the regularized
Bingham law~\eqref{binghamfluid} with the same plastic viscosity, $\mu_0 = \mu$.
In addition, for radii smaller than $\xi = 0.125$, we systematically found that
the optimal value of the coupling parameter $\beta$ equals $0$ also for
non-Newtonian fluids, leading to the the full-slip interface conditions at
$\Gamma$. This result is especially relevant in simulations of casting process
in concrete structures, where $\xi \le 0.125$ by construction requirements, see
\cite{CSNEN2}.
\subsubsection{Bingham flow}

This subsection illustrates the results for a Bingham fluid of plastic viscosity
$\mu_0 = 20$, yield stress of $\tau_0 = 20$, and regularization parameter of
$m=15$. The results are reported for RVEs with $\xi = 0.125$ with $\beta = 0$,
and $\xi = 0.25$ with $\beta = 3$, for both horizontal and skew flows. Again,
homogenized solution is validated against fully resolved solution from DNS.
\begin{figure}[h]
\centering
\begin{tabular}{cc}
\includegraphics[width=.425\textwidth]{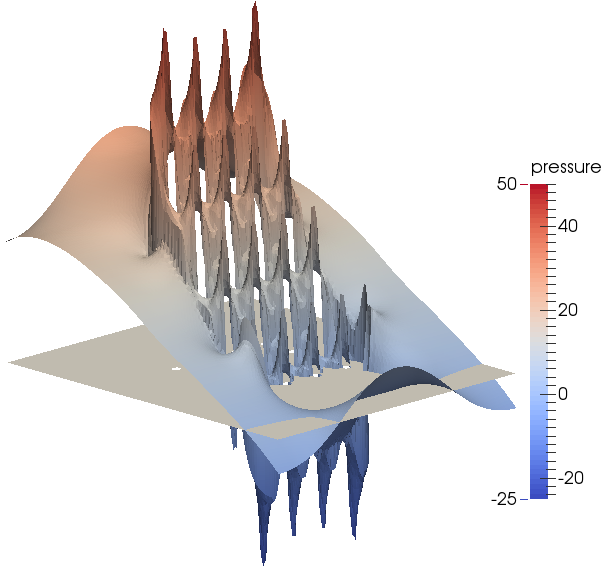} &
\includegraphics[width=.425\textwidth]{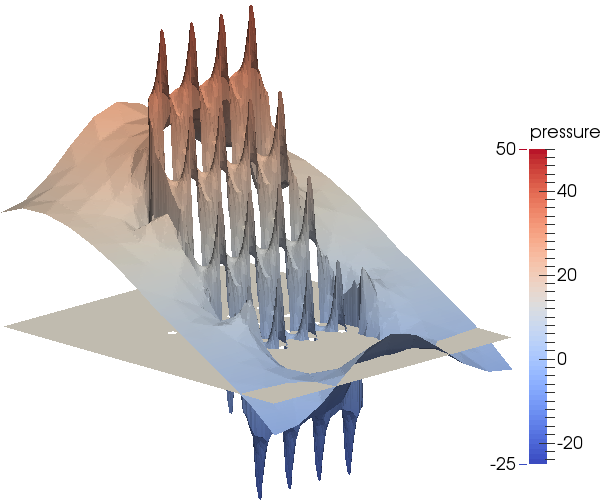} \\
\includegraphics[width=.45\textwidth]{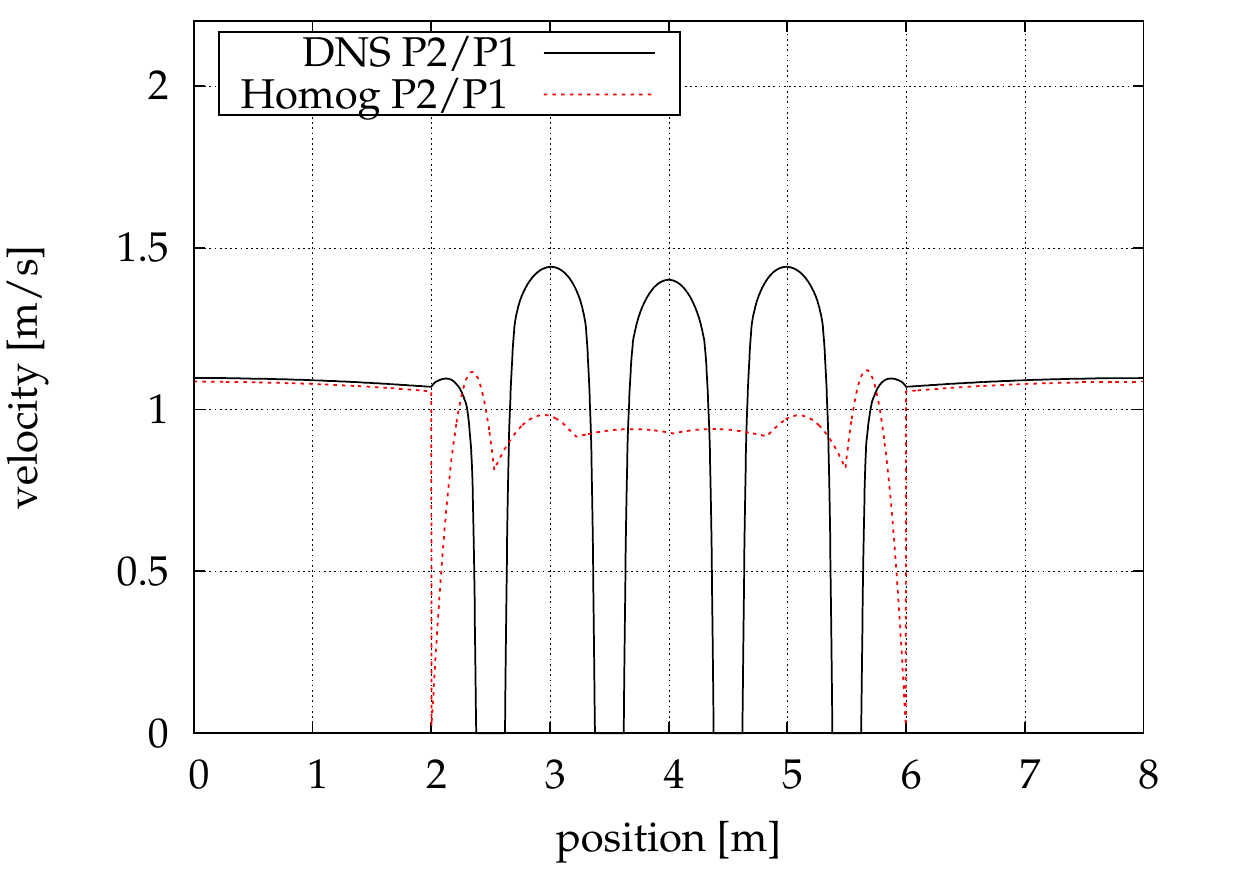} &
\includegraphics[width=.45\textwidth]{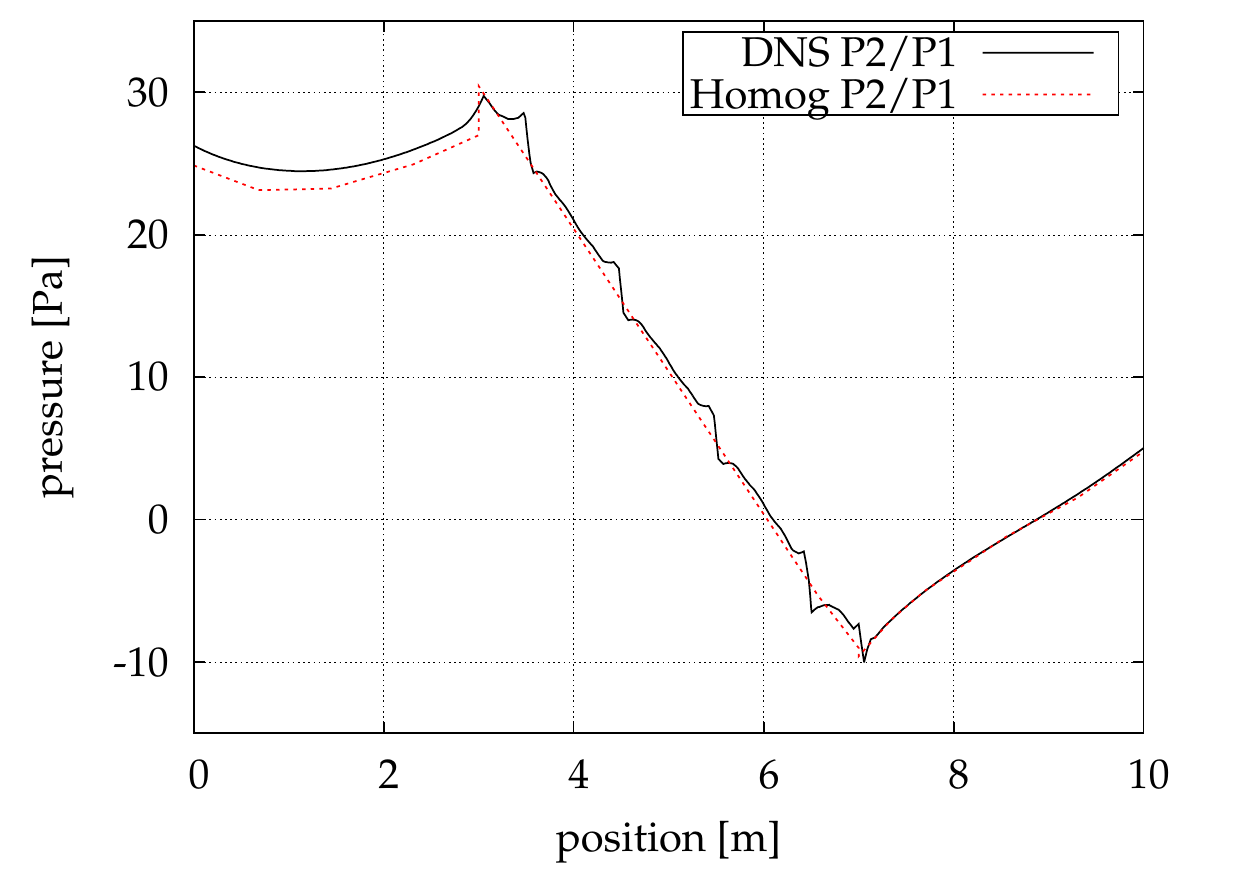} 
\end{tabular}
\caption{\textit{Flow over reinforced area: comparison of velocity and pressure
profiles for Bingham fluid with $\mu_0 = 20$, $\tau_0 = 20$, and $m = 15$, and
RVE with $\xi = 0.125$.}}\label{second_q_0.125_bing}
\end{figure}

\begin{figure}[p]
\vspace*{-10mm}
\centering
\begin{tabular}{cc}
\includegraphics[width=.425\textwidth]{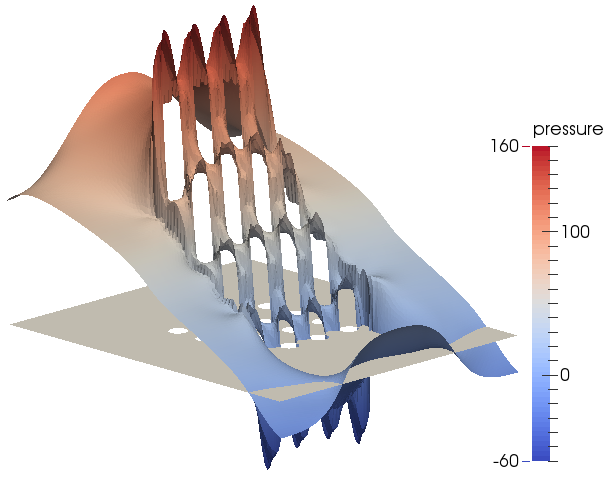} &
\includegraphics[width=.425\textwidth]{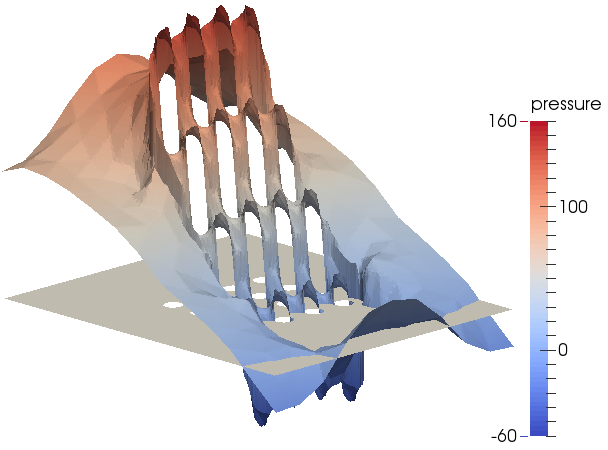} \\
\includegraphics[width=.45\textwidth]{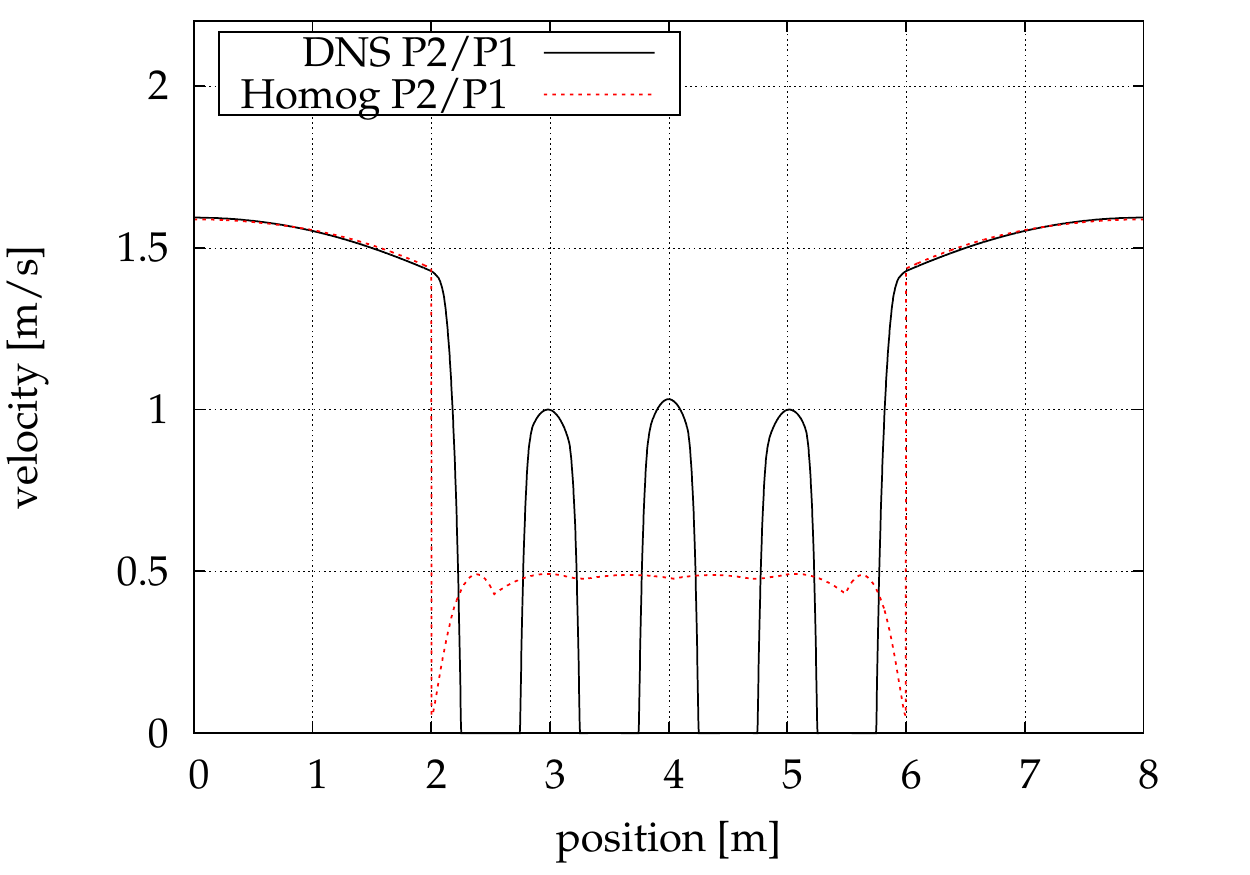} &
\includegraphics[width=.45\textwidth]{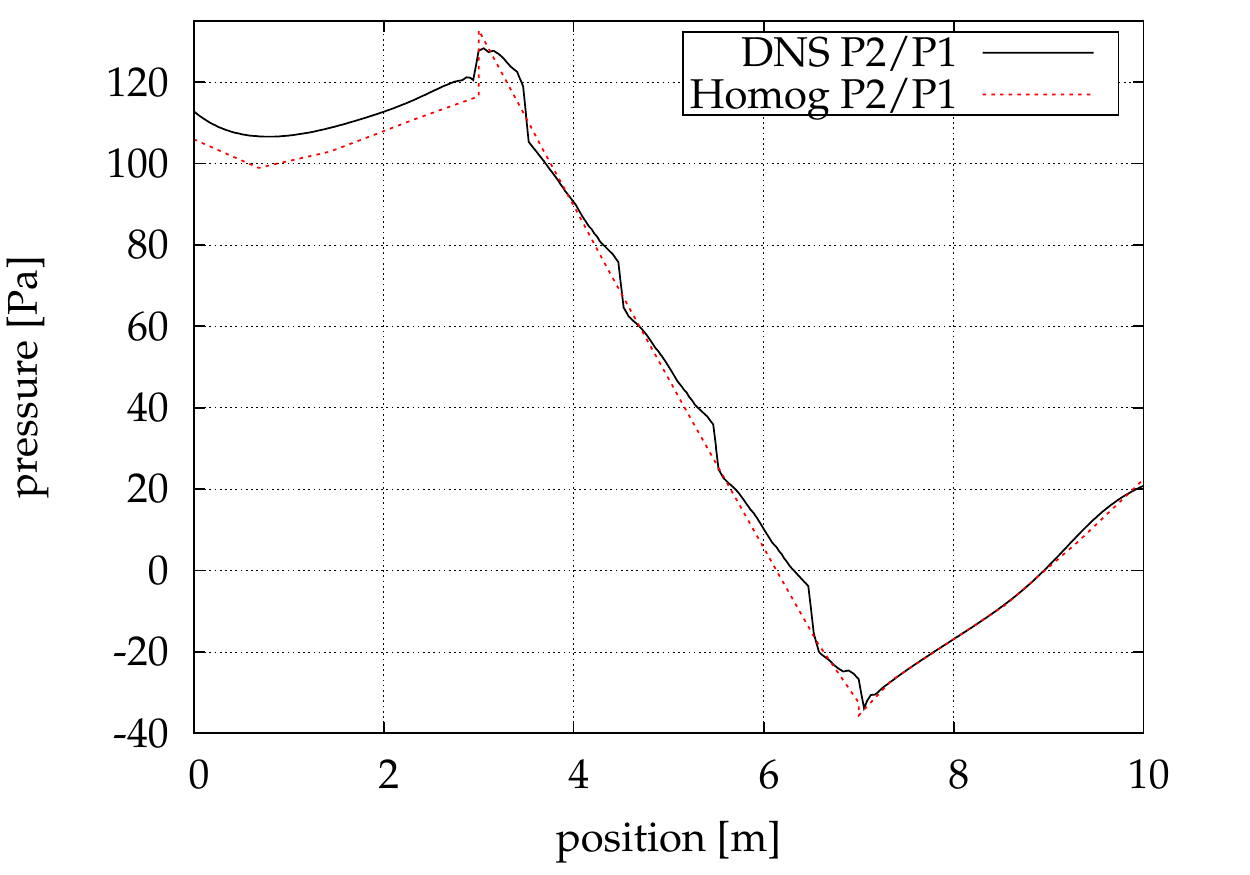} 
\end{tabular}
\caption{\textit{Flow over reinforced area: comparison of velocity and pressure
profiles for Bingham fluid with $\mu_0 = 20$ and $\tau_0 = 20$, $m = 15$.}}\label{second_q_0.25_bing}
\begin{tabular}{cc}
\includegraphics[width=.425\textwidth]{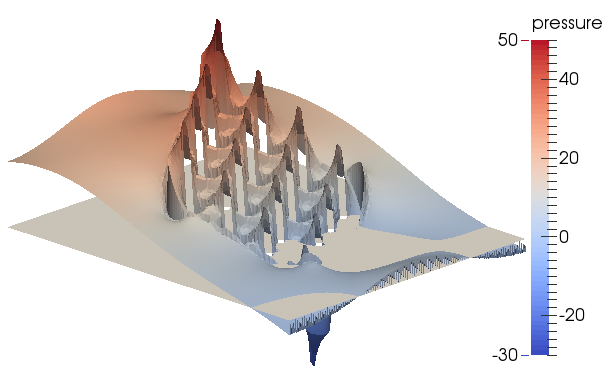} &
\includegraphics[width=.425\textwidth]{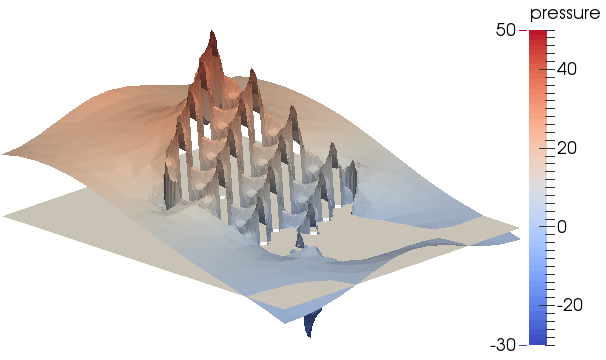} \\
\includegraphics[width=.45\textwidth]{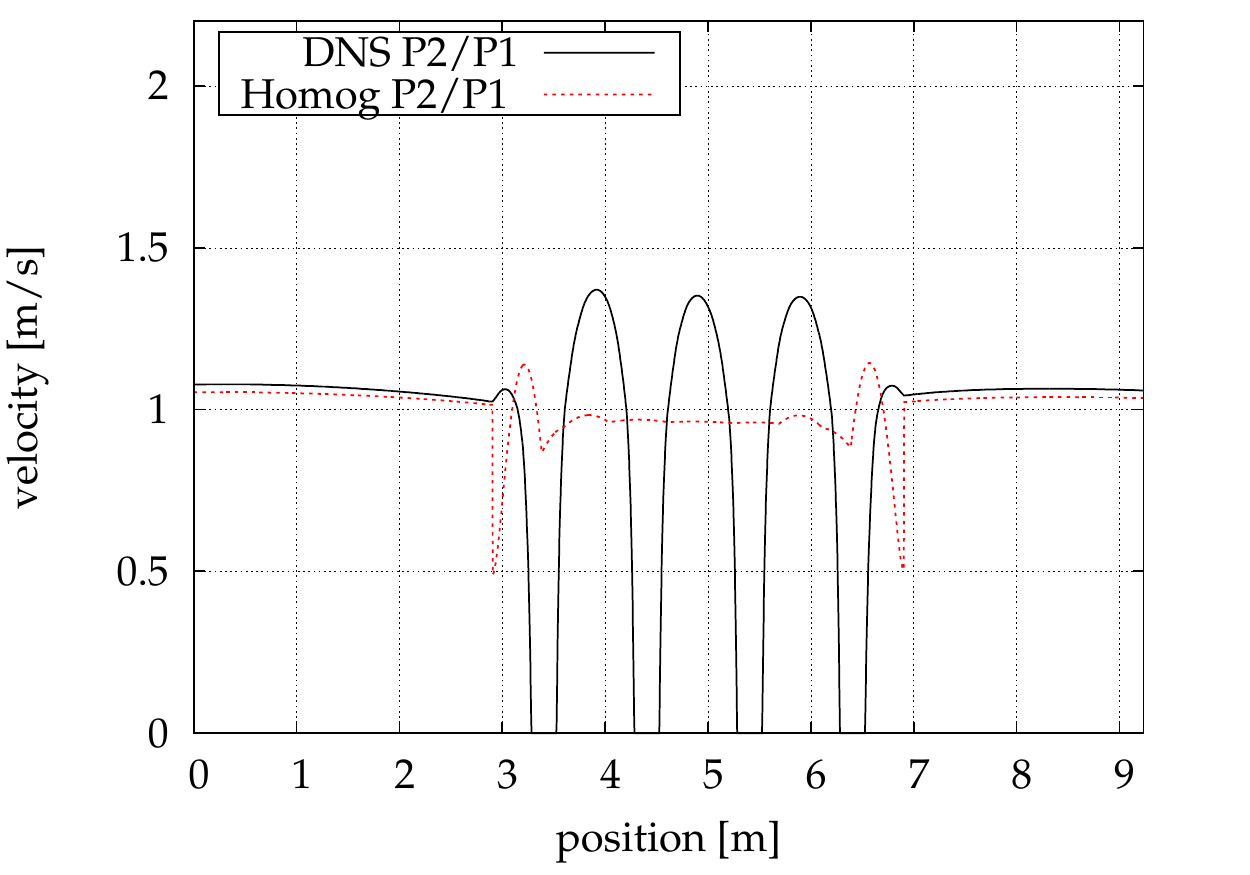} &
\includegraphics[width=.45\textwidth]{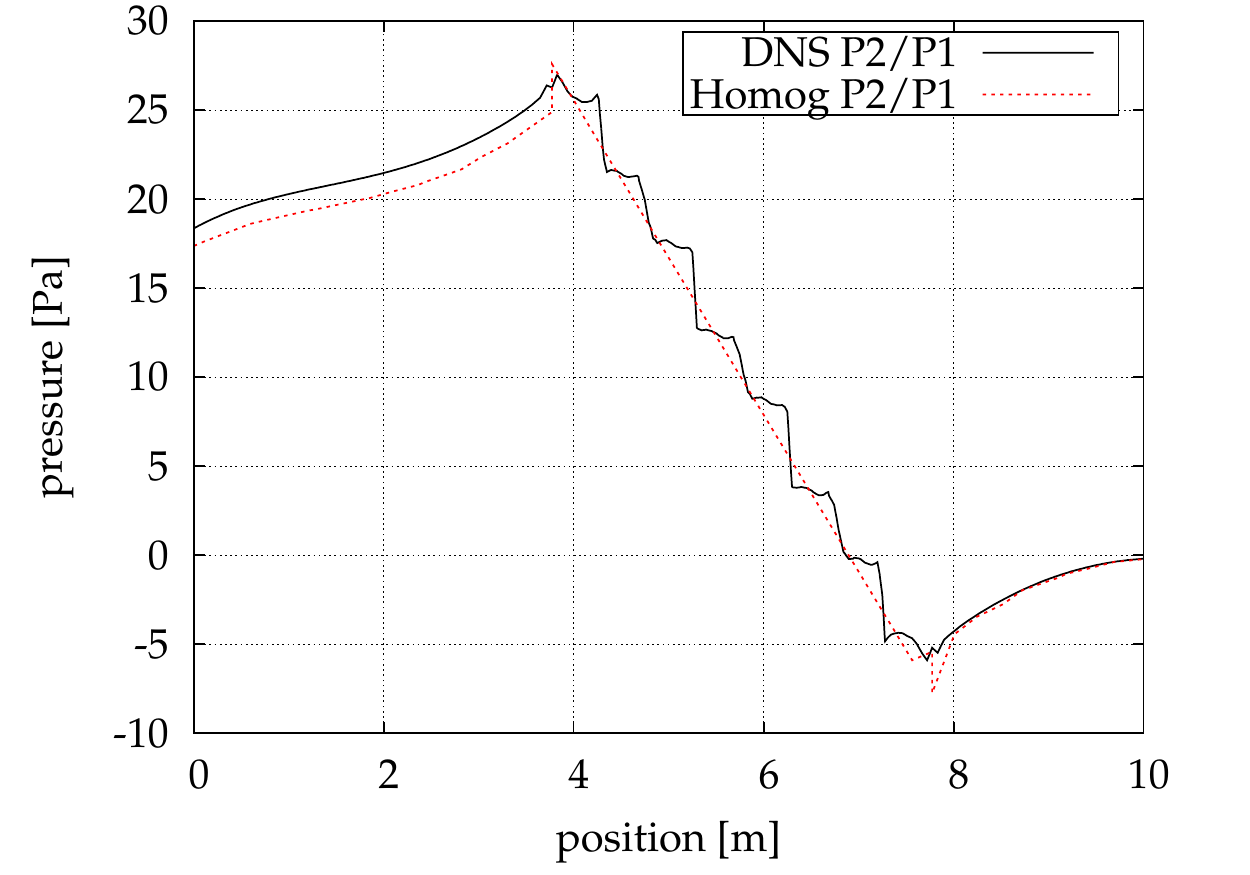} 
\end{tabular}
\caption{\textit{Flow over reinforced area: comparison of velocity and pressure
profiles for Bingham fluid in rotated setup with $\mu_0 = 20$, $\tau_0 = 20$,
$m = 15$, and RVE with $\xi = 0.125$.}}\label{second_q_0.125_bing_rot}
\end{figure}

Fig.~\ref{second_q_0.125_bing} shows the results for the horizontal setting and
for $\xi = 0.125$. The top row compares the pressure distributions of
homogenized and fully resolved solutions in axonometric view. Note that here we
show the reconstructed pressure with the added influence  of the sub-scale pressure
$p^\mathrm{S}$ obtained from the solution in the RVE according to the
expansion~\eqref{pressureExpansion}. The bottom row compares the velocity
profiles over the section through the third column of obstacles (on the left),
and the pressure profile over the horizontal section through the whole domain
(on the right). These plots show only the averaged pressure and velocity. The
results are in very good agreement, which can be seen in the bottom right
picture with the pressure along the section through the whole domain. The
macroscopic pressure gradient is captured extremely well. The oscillations close
to the obstacles of course cannot be captured via homogenization approach.
However, the average is in perfect agreement with the result obtained with DNS.
The $\max$ norm error in average velocity inside the reinforced area is again less than 1\%.
The results for $\xi = 0.25$, shown in Fig. \ref{second_q_0.25_bing}, support
the same conclusions.

\begin{figure}[h]
\centering
\begin{tabular}{cc}
\includegraphics[width=.425\textwidth]{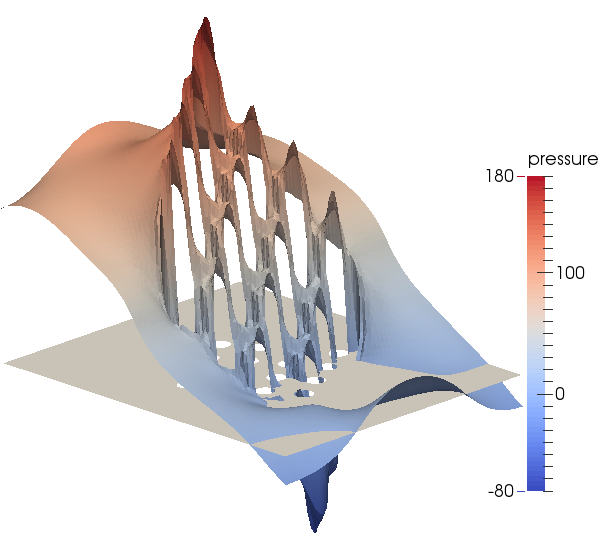} &
\includegraphics[width=.425\textwidth]{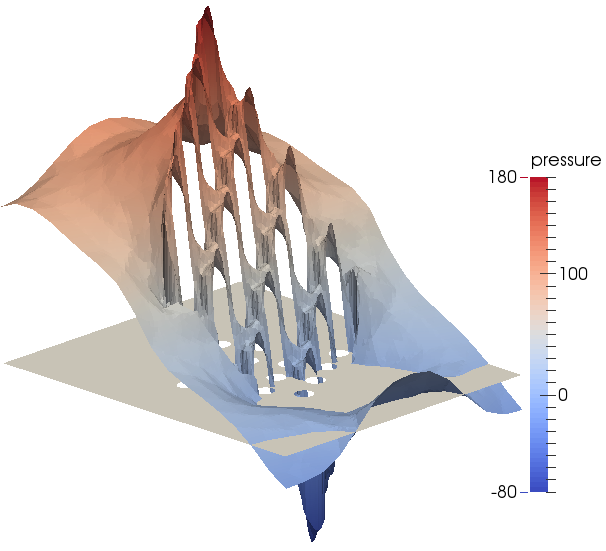} \\
\includegraphics[width=.45\textwidth]{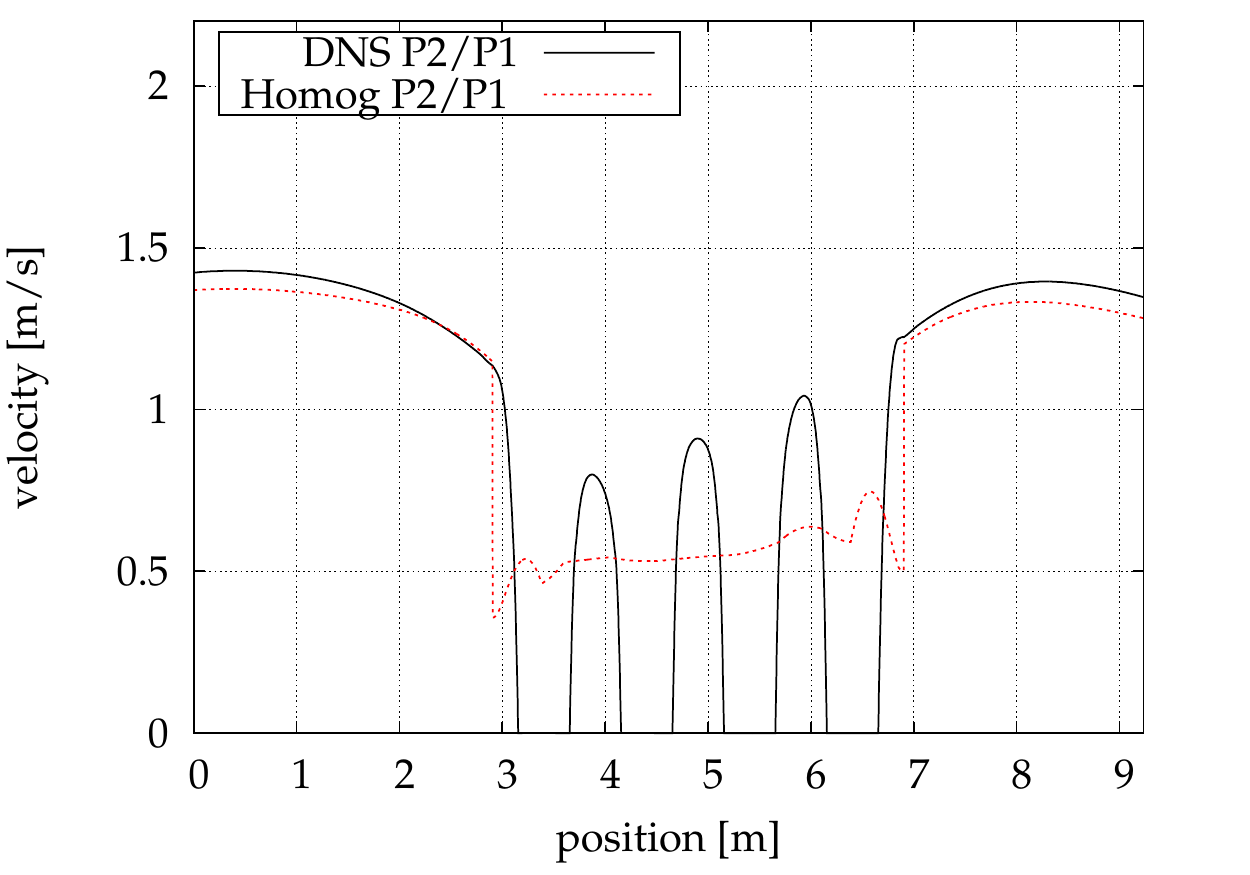} &
\includegraphics[width=.45\textwidth]{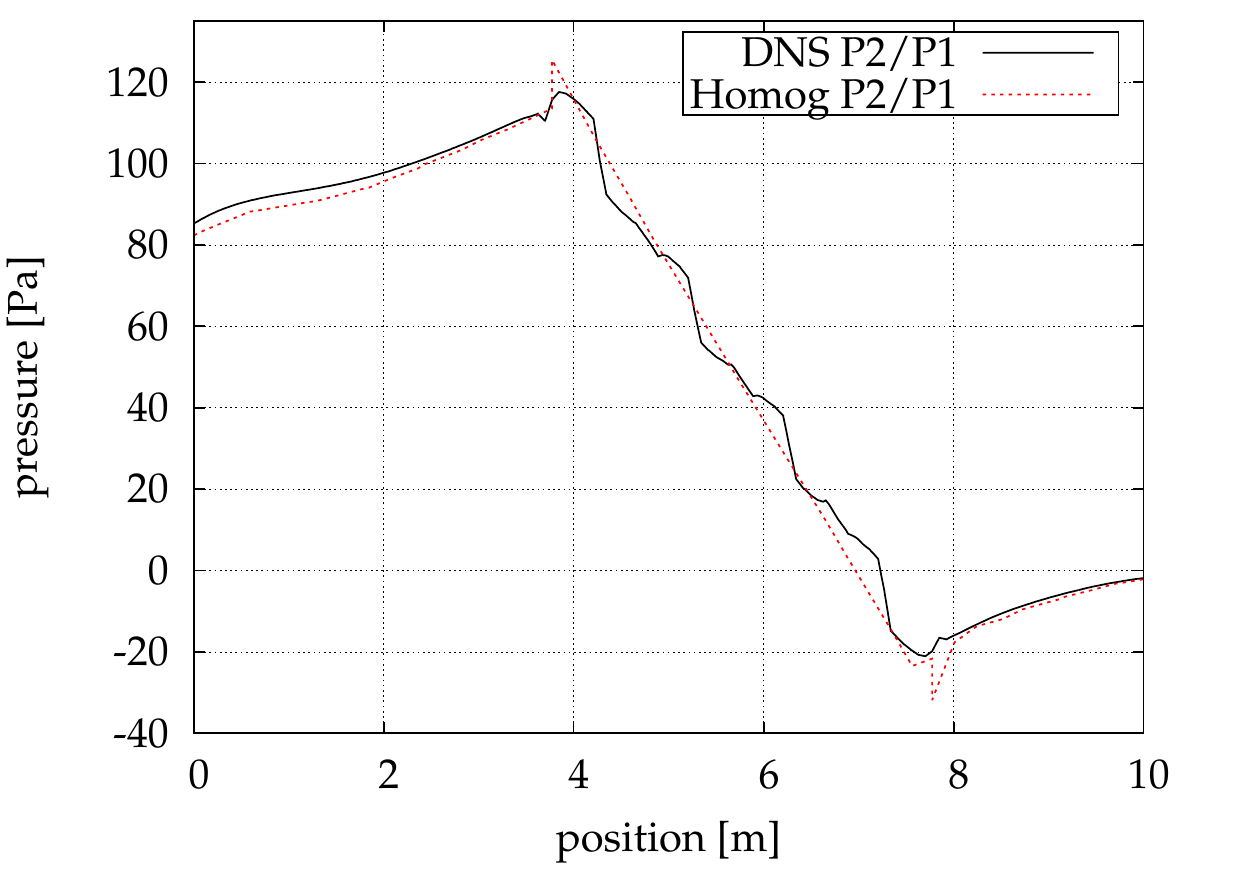} 
\end{tabular}
\caption{\textit{Flow over reinforced area: comparison of velocity and
pressure profiles for Bingham fluid in rotated setup with $\mu_0 = 20$, $\tau_0
= 20$, $m = 15$, and RVE with $\xi = 0.25$.}}\label{second_q_0.25_bing_rot}
\end{figure}

Figs.~\ref{second_q_0.125_bing_rot} and~\ref{second_q_0.25_bing_rot} show the
results obtained for the rotated setup. Although the flow pattern is
considerably more complex, the agreement of the results is also very good. The
error in the average velocity through the reinforced domain is less than $5\%$
in the $\max$ norm.
  
\section{Conclusions}\label{sec:conclusions}
  
In this paper, a computational homogenization approach to the flow of a
non-Newtonian fluid through a perforated domain has been developed, with
potential applications in simulations of concrete casting. We have presented a
unified formulation of a coupled Stokes-Darcy system obtained by a
variationally consistent homogenization of the Stokes flow in the porous sub-domain.
Specifically, by introducing the decomposition of the pressure and corresponding
test functions into macro and sub-scale parts in the perforated domain, the
problem splits into a Darcy-type flow on the macro-scale, coupled to a nonlinear
Stokes problem on the sub-scale, which is driven by the macroscopic pressure
gradient. The consistent linearization of the ensuing problem has been
presented, together with a numerical solution algorithm. The capabilities and
potentials of the developed methodology have been illustrated by several
examples comparing the homogenized solution with fully resolved simulations. On
the basis of the obtained results, we conjecture that

\begin{enumerate}

\item in the reinforced domain, the homogenized non-linear Darcy law allows for
a systematic and accurate incorporation of the effect of reinforcement into
homogeneous models of (not only) fresh concrete,

\item in simulations of casting of self-compacting concrete, the full-slip
conditions at the interface between the Stokes and Darcy domains deliver
sufficiently accurate results with respect to the fully resolved model,

\item the parameters of the Beavers-Jones-Saffmann constitutive law determined
by J\"{a}ger and Mikeli\'{c}~\cite{JagerMikelic2000} for a \emph{free} Newtonian
fluid flow over a porous bed do not reflect the coupling between the Stokes and
Darcy domains accurately, at least for the considered test cases. These findings seem to be consistent with a recent study of \emph{forced} filtration of a Newtonian fluid into a porous medium by Carraro et al.~\cite{Carraro:2015:EIC}, where different boundary conditions were derived by the refined asymptotic analysis and validated by direct numerical simulations. Additional studies are needed to clarify the interface conditions for non-Newtonian fluids.

\end{enumerate}

\section*{Acknowledgments}
The authors would like to acknowledge the support by the Czech Science
Foundation under project 13-23584S. In addition, Jan Zeman would like to thank
Dr. Manoj Kumar Yadav (Mahindra \'{E}cole Centrale) for many helpful discussions
on the works by J\"{a}ger, Mikeli\'{c}, and co-workers.


\end{document}